\documentclass[twocolumn,prb,superscriptaddress,showpacs]{revtex4}
\usepackage{amsfonts}
\usepackage{amssymb}
\usepackage{amsmath}
\usepackage{graphicx}
\usepackage{dcolumn}   
\usepackage{bm}        
\usepackage{flafter}
\usepackage{color}
\usepackage{hyperref}  
\usepackage{textcomp}

\usepackage{amsbsy}
\usepackage{units}
\usepackage{ulem}
\usepackage{verbatim}
\usepackage{placeins}

\newcommand{\STO}{SrTiO$_{3}$}
\newcommand{\LAO}{LaAlO$_{3}$}
\newcommand{\ttwog}{$t_{2g}$} 

\begin{document}

%
\title{Hydrostatic pressure response of an oxide two-dimensional electron system}
\date{\today}
 \author{J.~Zabaleta} 
 \affiliation{Max Planck Institute for Solid State Research, 70569 Stuttgart, Germany}
 \author{V.S.~Borisov}
 \affiliation{Institute of Theoretical Physics, Goethe University, 60438 Frankfurt am Main, Germany}
 \author{R.~Wanke}
 \affiliation{Max Planck Institute for Solid State Research, 70569 Stuttgart, Germany}
 \author{H.O.~Jeschke}
 \affiliation{Institute of Theoretical Physics, Goethe University, 60438 Frankfurt am Main, Germany}
 \author{S.C.~Parks,$^1$ B.~Baum,$^1$ A.~Teker,$^1$ T.~Harada,$^1$ K~Syassen}
 \affiliation{Max Planck Institute for Solid State Research, 70569 Stuttgart, Germany}
 \author{T.~Kopp}
 \affiliation{Center for Electronic Correlations and Magnetism, University of Augsburg, 86135 Augsburg, Germany}
 \author{N.~Pavlenko}
 \affiliation{Max Planck Institute for Solid State Research, 70569 Stuttgart, Germany}
 \affiliation{Center for Electronic Correlations and Magnetism, University of Augsburg, 86135 Augsburg, Germany}
 \author{R.~Valent\'i}
 \affiliation{Institute of Theoretical Physics, Goethe University, 60438 Frankfurt am Main, Germany}
 \author{J.~Mannhart}
 \affiliation{Max Planck Institute for Solid State Research, 70569 Stuttgart, Germany}

(Phys. Rev. B \textit{in press})
\begin{abstract}
Two-dimensional electron systems with fascinating properties exist in multilayers of standard semiconductors, on helium surfaces, and in oxides. Compared to the two-dimensional (2D) electron gases of semiconductors, the 2D electron systems in oxides are typically more strongly correlated and more sensitive to the microscopic structure of the hosting lattice. This sensitivity suggests that the oxide 2D systems are highly tunable by hydrostatic pressure. Here we explore the effects of hydrostatic pressure on the well-characterized 2D electron system formed at {\LAO}-{\STO} interfaces \cite{Ohtomo2004} and measure a pronounced, unexpected response. Pressure of $\sim$2 GPa reversibly doubles the 2D carrier density $n_{s}$ at 4 K. Along with the increase of $n_s$, the conductivity and mobility are reduced under pressure. First-principles pressure simulations reveal the same behavior of the carrier density and suggest a possible mechanism of the mobility reduction, based on the dielectric properties of both materials and their variation under external pressure.
\end{abstract}

\smallskip
\maketitle

\section{Introduction} Pressure is a powerful tool to study and tailor the
properties of solids (for reviews, see e.g. Refs.~\onlinecite{Paul1998,
Schilling2007}). In state-of-the-art field effect transistors, for example,
epitaxial strain is applied to enhance the electron mobility in silicon and
thereby to raise the switching speed. In quantum cascade lasers,
layer-straining is used to tailor the band structure for optimal
performance.~\cite{Paul1998} Hydrostatic pressure has furthermore been applied
to induce superconductivity in H-S compounds with reported transition
temperatures as high as 203 K.~\cite{Drozdov2015} Striking changes of the
electronic properties upon application of pressure have also been found in
oxides, which due to their structural degrees of freedom are intrinsically even
more sensitive to lattice strain than conventional semiconductors. To give but
a few examples, crystal fields, orbital occupancy, polarizations, and exchange
coupling in oxides are influenced substantially by slight distortions of ionic
sublattices. Pressure-induced changes of the structural and electronic
properties have been investigated extensively in bulk oxides and have been
demonstrated for thin films. By applying an epitaxial biaxial strain of
$\sim$0.9\,{\%}, for example, nominally paraelectric {\STO} films have been
turned into ferroelectrics with a $T_{c}$ of 293 K.~\cite{Haeni2004} The study
of the effects of epitaxial strain in {\LAO}-{\STO} heterostructures revealed
that the critical thickness of the {\LAO} layer required to generate the 2D
electron systems (2DES) and their carrier concentration depend on the epitaxial
strain of the {\STO}.~\cite{Bark2011, Nazir2014a, Nazir2014b} To open a new
route for the exploration of the electronic properties of complex oxides under
tunable uniaxial and biaxial stress, a novel piezo-based technique is currently
being developed.~\cite{Hicks2014}

Uniaxial and biaxial stress change the shapes of the unit cells but alter their volumes only little. For this reason, they do not provide an option to explore and benefit from those electronic effects that are induced by volume changes of the unit cell. These changes are accessible and have been explored by applying hydrostatic pressure. For example, in cuprate high-$T_{c}$ superconductors the record $T_{c}$ of 164 K was induced by a pressure of 15~GPa,\cite{Schilling2007, Gao1994} which enhanced the charge transfer between the CuO$_{2}$ layers and off-layer structural units. More recently, the Curie temperature of SrRuO$_{3}$ films was reduced from ~150 to 77 K by applying hydrostatic pressure up to 23\,GPa.~\cite{Lemarrec2002} In bulk n-doped {\STO} a non-monotonic dependence of carrier density and mobility was measured in pressure ranges $\leq$1~GPa.~\cite{Laukhin2012} Whereas the above-mentioned studies of the effects of hydrostatic pressure on oxides focused on the properties of three-dimensional electron systems, we study here the pressure effects on a 2DES. 

The 2DES at the interface between {\LAO} and {\STO}, two band insulators, has
been widely investigated since its discovery in 2004.~\cite{Ohtomo2004} It has
served as a model system to study the origin of conducting layers at oxide
interfaces and has shown a rich phenomenology, including the coexistence of
magnetism and superconductivity (for an overview, see, e.g.
Ref.~\onlinecite{Hilgenkamp2013}). A polar discontinuity develops at the
interface between TiO$_{2}$-terminated, (001)-oriented {\STO} and {\LAO}. For
{\LAO} layers with a thickness exceeding 3 unit cells (u.c.),~\cite{Thiel2006}
the resulting electric field moves charge carriers from the surface of the
{\LAO} to the interface, where an n-type 2DES is formed for the TiO$_{2}$
termination of {\STO}. The formation of the 2DES and its properties can be
influenced by the presence of defects such as oxygen vacancies (for an
overview, see e.g. Refs.~\onlinecite{Huijben2009, Wadati2009,Pavlenko2012}). At
ambient pressure the electrons at the interface occupy the {\ttwog} Ti 3$d$
levels, split by 50~meV into low-energy $d_{xy}$ and higher-energy $d_{xz}$ and
$d_{yz}$ bands.~\cite{Popovic2008,Salluzzo2009}

\begin{figure}
\includegraphics[width=8.8cm]{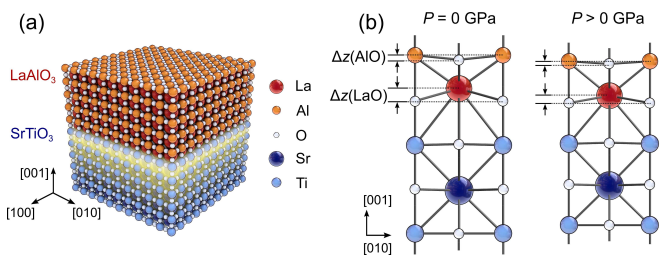}
\caption{Illustration of the atomic structure of the {\LAO}-{\STO} interface. Panel (a) shows a sketch of the heterostructure with the 2DES at the interface marked in yellow. Panel (b) illustrates the displacements $\Delta z$(LaO) of the La$^{3+}$ and O$^{2-}$ ions in the LaO sublayer and of the Al$^{3+}$ and O$^{2-}$ ions in the AlO$_{2}$ sublayer, $\Delta z$(AlO). Hydrostatic pressure reduces $\Delta z$(LaO) and $\Delta z$(AlO). The much smaller ionic displacements in the {\STO} are not indicated.}
\label{FigLAOSTO}
\end{figure}

Pressure acting on perovskite-related oxides compresses bonds and modifies the
internal structural parameters. For the {\LAO}-{\STO} bilayers, the hydrostatic
pressure superimposes the stress resulting from the lattice mismatch of
$\sim$3\,{\%} and decreases it. The influence of epitaxial strain on the 2DES
was addressed in first-principles model calculations of surface-free
heterostructures by Nazir \textit{et al.}\cite{Nazir2014a, Nazir2014b} and
experimentally by Bark and collaborators.\cite{Bark2011} The latter group found
that while the carrier concentration is reduced in compressively-strained
{\STO} as a result of dipole formation within the {\STO}, no 2DES is formed on
tensile-strained {\STO}. These results immediately lead to the question of how
the interface 2DES is modified by hydrostatic pressure. Owing to the unit-cell
compression it may significantly differ from the response to epitaxial strain.

In this work we combine experimental observations with first-principles electronic 
structure calculations to study the effect of pressure on the microstructure of the
{\LAO}-{\STO} interface and its implications on the electronic properties of the 2DES 
(see Fig.~\ref{FigLAOSTO}). We find that the lattice reconstruction in the {\LAO} film, which
acts against the interface charge, is partially suppressed by the external
pressure thereby reducing the lattice screening of the polar discontinuity and
enhancing charge transfer towards the oxide interface. This mechanism sheds
light on the reversible increase of the carrier concentration of the 2DES
measured in transport experiments up to $\sim$2 GPa. This
observation has been also confirmed
in recent independent  experimental work by D. Fuchs \textit{et al.}.~\cite{Fuchs2015}
Furthermore, our calculations of
static dielectric properties in  {\LAO} and {\STO} under pressure indicate a
reduction of carrier mobility as observed in our measurements.
The behavior of this oxide system is opposite in sign to the pressure response of semiconductor quantum wells,~\cite{Ernst1994} where the carrier concentration is found to decrease with increasing pressure (see Fig.~\ref{Fig_AlGaAs} in the Supplement). Our findings reveal that, in contrast to biaxial
strain, hydrostatic pressure is a suitable tool for enhancing
the carrier density of the 2DES in this oxide system.

\section{Methods}
\subsection{Sample growth, pressure cells and transport measurements}
\begin{figure}[tbp]
\includegraphics[width=8.8cm]{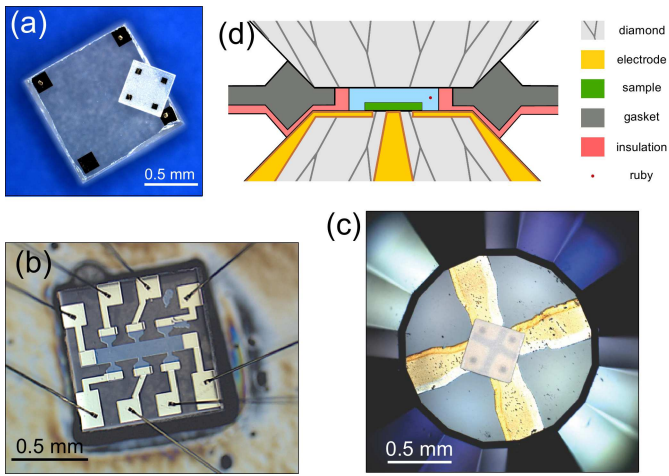}
\caption{(a) Optical microscopy image showing samples for both a diamond anvil cell (small sample) and for the piston cylinder cell (large sample). Both samples are in a van-der-Pauw configuration. (b) Optical micrograph of a sample with a patterned and contacted Hall bar. The sample is 1\:$\times$\:1\:$\times$0.2\:mm$^{3}$ in size and the Hall bar is 100~$\mu$m wide. (c) Optical micrograph of a sample placed on the bottom diamond of the diamond anvil cell patterned with the Ti/Pt/Au trilayer. The four metallic contacts of the sample (visible through the translucent substrate) touch the four electrodes of the diamond tip. (d) Sketch of the diamond-anvil cell setup. The gasket hole has a diameter of 0.85-0.9~mm. Ti/Pt/Au electrodes evaporated on the lower diamond serve as leads to contact the sample.} 
\label{FigExp}
\end{figure}

Samples were grown by pulsed laser deposition from single-crystal {\LAO} targets onto TiO$_{2}$-terminated {\STO} single-crystal substrates at 800$^\circ$C and 8\;$\times$\;10$^{-5}$\;mbar O$_{2}$ partial pressure followed by two annealing steps at 600$^\circ$C and 400$^\circ$C for 1\;h at 400\;mbar of O$_{2}$ partial pressure. The thickness of the {\LAO} layer was monitored by in-situ reflective high-energy electron diffraction. Samples were patterned into van der Pauw configurations and into 100 $\mu$m-wide Hall bridges by optical lithography and Ar ion sputtering, with electron-beam evaporation used to deposit Ti-Au contacts into the buried electron system (Fig.~\ref{FigExp}(a),(b)). Patterning and contacting was done as described in Ref.~\onlinecite{Schneider2006}, except for using 50 nm of polycrystalline and 6 unit cells of epitaxial {\LAO}. Hydrostatic pressure was applied using either a diamond anvil cell or a piston cylinder cell. In the diamond anvil cell transport measurements the heterostructure samples are encased between the small tips of the diamonds. Therefore, several series of {\LAO}-{\STO} samples were fabricated and cut to typical sizes of 350\:$\times$\:350\:$\times$65\:$\mu$m$^{3}$ (Fig.~\ref{FigExp}(a),(c)). As the piston cylinder cell allows larger samples, the typical size of the samples used for the piston cell measurements was the 1\:$\times$\:1\:$\times$0.2\:mm$^{3}$ (Fig.~\ref{FigExp}(a),(b)). Control measurements confirmed that the sample properties were not affected by the cutting (see Supplement Fig.~\ref{FigBondedSample}).

As illustrated in Fig.~\ref{FigExp}(d), we used diamonds of 1.25 and 1.5~mm tip diameters. Inconel gaskets (250~$\mu$m thick) were pre-pressed, drilled, and then insulated with a mixture of Stycast 1266 and Al$_{2}$O$_{3}$ powder. The gasket hole had a diameter of between 850-900~$\mu$m and an initial height of 170-200 $\mu$m. The cleaned diamond surface of the lower anvil was patterned with a metallic trilayer (50 nm Ti/50 nm Pt/500 nm Au) by electron-beam evaporation through a stainless-steel shadow mask. Contacts were annealed in vacuum at 650$^\circ$C for 20~min. Daphne 7373 oil was used as pressure medium; the pressure was measured using ruby luminescence. Measurements with a piston cylinder cell (Almax Easy Labs) were also performed using a manganin coil as pressure sensor and Daphne 7373 as pressure medium.

To avoid photo-excited carriers, all transport measurements were performed in darkness, after keeping the samples dark for at least 12~h. Prior to the application of pressure, all samples were characterized by transport measurements. Measurements were carried out in the van der Pauw configuration with lead permutations, and in a Hall bar configuration. Measurement currents were 1-5~$\mu$A at room temperature and 10~$\mu$A at low temperature for both the van der Pauw samples and the Hall bridges. The power dissipated at the current leads was less than 2 $\mu$W. Carrier density values were extracted from the Hall signals taken with decreasing and increasing temperature. 
For the diamond-anvil-cell measurements, the pressure was increased in steps of $\sim$0.2~GPa, followed by a 1-2~h waiting time for the pressure and its effects to become steady. Temperature, pressure, and resistance were constantly monitored to follow the stabilization process, which we found to depend on the size of the pressure change and on the sample. For the piston-cylinder-cell measurements, the pressure was increased at room temperature with an oil press in steps of $\sim$2~MPa. The pressure was monitored via the resistance change of a manganin wire inside the sample chamber. At the desired pressure values, the piston cylinder cell was mounted in a physical properties measurement system (PPMS by Quantum Design) for the transport measurements.

\subsection{First principles calculations}
\begin{figure*}
  \includegraphics[width = 0.87\textwidth]{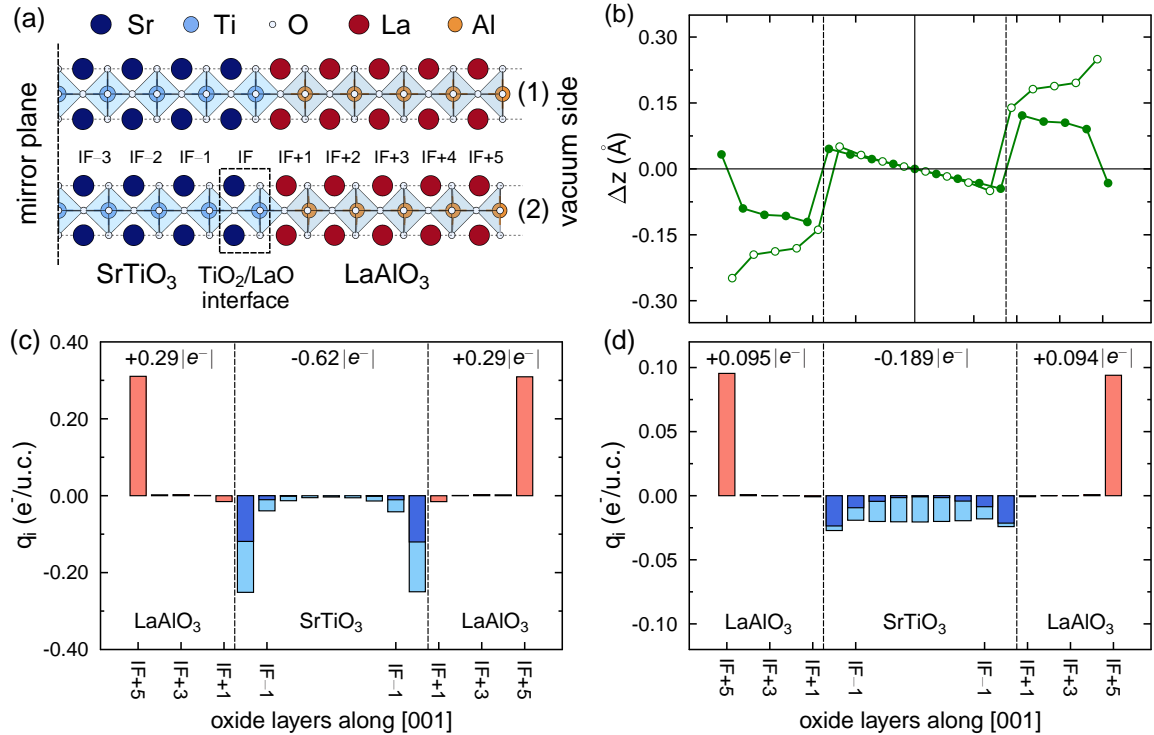}
  \caption{(a) Half of the (\LAO)$_n$/(\STO)$_m$/(\LAO)$_n$ supercell
  ($n=5$, $m=8.5$) representing the ideal (panel (1)) and fully relaxed (panel (2))
  TiO$_2$/LaO-terminated n-type interface (vacuum side is not fully shown here).
  The interface unit cell of {\STO} is labeled by IF, and the others by (IF$\pm$1),
  (IF$\pm$2) etc. Structure figures have been produced with \textsc{vesta3}.\cite{Momma11} 
  (b) Ionic displacements in \textit{A}O
  ($A={\rm Sr}$, La; open symbols) and \textit{B}O$_2$ ($B={\rm Ti}$, Al; filled
  symbols) oxide layers of the (\LAO)$_5$/(\STO)$_{8.5}$/(\LAO)$_5$ slab relaxed
  internally with fixed $ a_\perp = 3.905\:\mathrm{\AA} $. 
  Positive values on the right of the figure correspond to La and Al cations shifting above 
  the oxygen ions, i.e. closer to the vacuum on the corresponding side of the slab. 
  Due to the imposed mirror symmetry of the supercell, the ionic displacements on the 
  left side of the structure have the opposite signs compared to the right side. 
  (c,d) Unit-cell resolved charge distribution in
  the (\LAO)$_5$/(\STO)$_{8.5}$/(\LAO)$_5$ slab for the  (c) unrelaxed  and 
 (d) internally
  relaxed  structures with $ a_\perp = 3.905\:\mathrm{\AA} $. 
  Each bar on the plots represents an oxide layer unit cell. ``Negative'' values correspond 
  to n-type carriers (electrons in the conduction band of {\STO}) and ``positive'' values to 
  p-type carriers (holes in the valence band of {\LAO}). In the {\STO} film, the darker blue 
  bars show the contribution of the $d_{xy}$ states, compared to the total local charge 
  (light-blue bars). The charge summed over all layers of {\STO} and both {\LAO} films is 
  indicated on top of each panel.}
  \label{f:lat-reconstr}
\end{figure*}

To study the n-type \LAO-\STO~(001) interface in the thin film geometry, 
we constructed (\LAO)$_n$/(\STO)$_m$/(\LAO)$_n$ supercells with two symmetric
TiO$_2$/LaO-terminated interfaces and different number of oxide unit
cells along the [001] axis ($n=3-6$ and $m=8.5,\:20.5$) and 
($1 \times 1$) in-plane dimensions. 
Calculations for ($\sqrt 2 \times \sqrt 2$) 
in-plane dimensions were also performed for some test cases. 
The case of $n=5$ and $m=8.5$ is presented in Fig.~\ref{f:lat-reconstr}(a) 
where half of the supercell is depicted. The other half is obtained using 
the mirror symmetry with respect to the center of the {\STO} slab 
(mirror plane).  
We considered the stoichiometric case with no defects or vacancies.
 A vacuum layer of more than 25~\AA{} was inserted to prevent any spurious
interactions between the periodic images as in previous slabs 
calculations.~\cite{Altmeyer2015,Borisov2015} The imposed mirror symmetry
of the supercell reduces the effect of a build-up dipole due to the
formation of the 2DES at both interfaces. 
 $P=0$ internal coordinates and lattice vector relaxations were done using a
$\Gamma$-centered ($8\times 8\times 1$) Monkhorst-Pack \textbf{k}-mesh which
provided atomic positions converged within $1.5\times 10^{-3}\:\mathrm{\AA}$.
Ionic forces in the relaxed structures were less than 25~meV/\AA{}. 

Electronic properties of
this system were calculated using density functional theory and
projector-augmented wave basis\cite{Bloechl1994a} as implemented in
VASP.\cite{Kresse1996,Hafner2008} Electronic wavefunctions were
expanded into plane waves up to the cutoff energy of
450~eV. The generalized-gradient approximation\cite{Perdew1996} to the
exchange-correlation energy was applied in combination with the GGA+$U$
scheme\cite{Dudarev1998} to treat electronic correlations for strongly
localized $3d$ and $4f$ states. For this purpose, we chose $U_{\rm eff} =
4\:\mathrm{eV}$ for the Ti $3d$ and $U_{\rm eff} = 8\:\mathrm{eV}$ for the
La $4f$ orbitals, which is consistent with the values often stated in the
literature. The latter values were used in order to place the La $4f$
empty orbitals at the correct energies.~\cite{Czyzyk1994}
The electronic density of states was calculated using the
tetrahedron method\cite{Bloechl1994b} on a finer ($16\times 16\times 2$) mesh. Layer-resolved local carrier densities are calculated by integrating the site-projected densities of states in each oxide layer.

\section{Zero-pressure Simulations}
 Since the 2DES is extremely sensitive to lattice relaxations
---a phenomenon seldom discussed in previous works---
 we first performed structural optimizations at $P=0$ by
relaxing all internal coordinates and the lattice vectors. The angles between
the lattice vectors remain 90\textdegree{} due to the lattice symmetry. 
For ($1 \times 1$) in-plane dimensions of the supercell, only
displacements along the [001] axis are allowed. 
We observe that the relaxation procedure reveals a significant reconstruction of
the lattice and of the electronic properties already at the level of relaxation
of atomic coordinates along [001] only (without volume relaxation), as can be
observed in Fig.~\ref{f:lat-reconstr}(b,d). These figures show results for the
(\LAO)$_n$/(\STO)$_m$/(\LAO)$_n$ supercell ($n=5$, $m=8.5$)
with fixed in-plane lattice parameter $ a_\perp =
3.905\:\mathrm{\AA} $.  

The unrelaxed slab displays a  2DES strongly
localized at the interface (Fig.~\ref{f:lat-reconstr}(c)), showing almost equal
occupations of  Ti $d_{xy}$ and $d_{xz}/d_{yz}$ orbitals. Their carrier density is large, approximately 0.3~$e^-$/u.c..
Optimization of the internal atomic positions changes this picture
drastically (Fig.~\ref{f:lat-reconstr}(d)). The observed shift of electronic states in {\LAO} becomes
smaller, 0.35~eV per unit cell (Fig.~\ref{f:DOS-layers}(b)).
Simultaneously, the charge density at the interface is considerably
reduced to 0.1~$e^-$/u.c., much less than the expected charge
transfer of 0.5~$e^-$/u.c. according to the polar catastrophe picture in its most simplified version.
As a matter of fact, there is a large amount of both theoretical and
experimental studies in the literature that argue about the origin of
this discrepancy (for a review, see Ref.~\onlinecite{Hanghui2010}).
Possible explanations include vacancy formation, cation disorder and lattice
reconstruction. In this paper, we report mainly on this last
mechanism, especially in view of our experimental findings regarding pressure
effects where we find similar carrier densities $\sim$\,0.1--0.2~$e^-$/u.c.
for the 2DES, also under pressure. We also find that after relaxation, the {\LAO} film
becomes distorted compared to
the perfect cubic structure, and adopts a tetragonal symmetry,
in agreement with previous studies.~\cite{Pentcheva2009,Pavlenko2011}
This happens because of the lattice mismatch between {\LAO} and {\STO}
which creates, already at ambient pressure, a strong tensile strain
in the {\LAO} film. More importantly, positively charged ions (La and Al)
are shifted by $\approx$\,0.2--0.3~\AA{} relative to the negative oxygen ions
which is reminiscent of a typical displacive ferroelectric transition.
The same picture holds for the {\STO} substrate where the displacements
$\Delta z$, defined as $ \Delta z = z_\mathrm{cation} - z_\mathrm{oxygen} $,
have the opposite direction and a much lower value. The distribution
of the ionic displacements in the whole heterostructure is shown in
Fig.~\ref{f:lat-reconstr}(b) (see Fig.~\ref{FigLAOSTO}(b) for a sketch of the interface).

In general, such a reconstruction pattern screens the interface
charge that builds the 2DES. This can be concluded from the sign of
the $\Delta z$ in {\LAO} and {\STO} (see Fig.~\ref{f:lat-reconstr}(b)).
In the {\LAO} film, the dipole moment related to the ionic displacements
creates a depolarizing electric field directed towards the interface,
the same as in the {\STO} substrate. This pushes electrons away from the interface and reduces
the accumulated charge. How this becomes important in the case of
non-zero external pressure will be discussed further below.

\begin{figure}
  \centering
  \includegraphics[width = 0.48\textwidth]{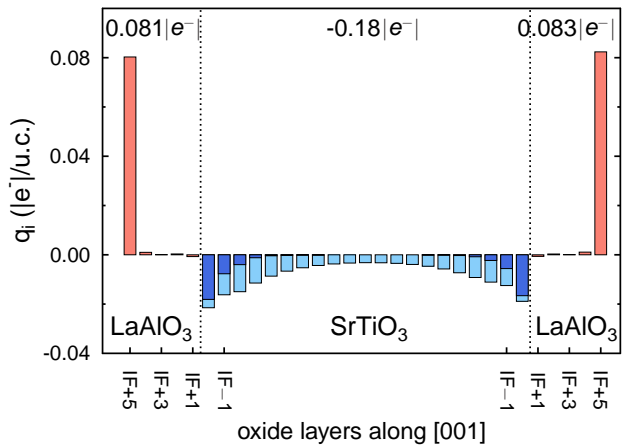}
  \caption{ Charge distribution in the fully relaxed (\LAO)$_5$/(\STO)$_{20.5}$/(\LAO)$_5$ slab 
    at zero pressure. Contribution of the $d_{xy}$ states is shown by the dark-blue bars. 
    }
  \label{f:Charge-distribution-20.5}
\end{figure}

Strictly speaking, the aforementioned internally relaxed structure
does not exactly simulate the zero-pressure case because of the mismatch between
the theoretical and experimental values of the in-plane lattice parameter.
In view of the sensitivity of the electronic structure to the possible lattice
and atomic reconstructions, we fully relaxed both the lattice vectors
and the internal coordinates which leads to a change of the in-plane lattice 
parameter $a_\perp$ from 3.905~\AA{} to 3.891~\AA{}.\footnote{One should note here 
that the resulting lattice constant depends on the choice of $U_{\rm eff}$ parameters 
in the GGA+$U$ approach. Pure GGA calculation ($U_{\rm eff} = 0$) gives the equilibrium 
$a_\perp$ value of 3.864~\AA{}.} This slight contraction of the lattice causes only 
a small reduction of the 2DES density. The orbital character and the overall localization 
of the relevant states remain unchanged. The main features of the metal-insulator 
transition in this system vs. the {\LAO} thickness are the same for the internally and 
fully relaxed slabs (Fig.~\ref{f:2DES-density-thickness}). These simulations reproduce 
the formation of the 2DES starting from 4 {\LAO} unit cells. Above the critical thickness, 
the 2DES density grows continuously with the thickness of the {\LAO} film, in agreement 
with the polar catastrophe scenario.
We also investigated the effect of oxygen octahedra tiltings
using supercells with $\sqrt 2 \times \sqrt 2$ in-plane dimensions. Fully
relaxed $\sqrt 2 \times \sqrt 2$
 structures with 3 and 5 unit cells of {\LAO} were compared at
zero pressure to the structures without tilting. These simulations show that
the LAO/STO interface with 3 {\LAO} u.c. remains insulating in the presence of
octahedra tiltings and the  
heterostructure with 5 {\LAO} u.c. is metallic as before.

\begin{figure}
\includegraphics[width=6.2cm]{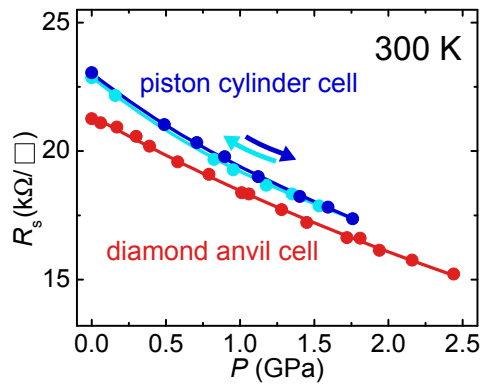}
\caption{Sheet resistance of the 2DES in a {\LAO}-{\STO} heterostructure measured at room temperature as a function of pressure. The {\LAO} layer of these samples is 5 u.c. thick. Lines serve as a guide to the eye. Experimental errors fall within the symbol size.}
\label{FigRevRep}
\end{figure}

One last note regarding the $P=0$ results is that the lattice reconstruction 
enhances the weight of the $d_{xy}$ states at the interface and that of the
$d_{xz}/d_{yz}$ states towards the middle of the {\STO} slab. 
This leads to a delocalization of the 2DES which spans  the whole {\STO} film 
with 8.5 u.c. (Fig.~\ref{f:lat-reconstr}). Considering larger slabs with 20.5~u.c. 
of {\STO} (Fig.~\ref{f:Charge-distribution-20.5}) clearly reduces the occupation
of the Ti $d_{xz}/d_{yz}$ orbitals in the middle of \STO. The 2DES at the interface is again composed 
mainly of the Ti $d_{xy}$ orbitals.
Our results indicate that at least 7~u.c. of {\STO} on each side of the
heterostructure are necessary to accommodate the quasi-2D electronic
states and, at the same time, leave a bulk-like region in {\STO}. 
The same applies to the structural distortions, which decay in 
{\STO} towards the middle of the slab (Fig.~\ref{f:Delta-z-20p5}), but, 
in general, show similar features observed in the smaller slabs. In addition, the fully relaxed heterostructure with 20.5 {\STO} u.c. has a larger in-plane lattice constant (3.92 \AA) than the slab with only 8.5 {\STO} u.c. (3.89 \AA). This shows that the in-plane lattice parameter in these simulations would approach the theoretical estimate for the bulk material (3.94 {\AA} in Ref.~\onlinecite{Piskunov2004}) in the limit of a very thick {\STO} film.
More importantly, we find that the integrated charge of the slab with 20.5~u.c. 
of {\STO} is already well reproduced by the  smaller supercell with 8.5~u.c. of 
{\STO}. This gives us confidence in performing the pressure simulations with the 
computationally less costly smaller slabs.

\section{Effect of pressure}
\subsection{Experimental results}
The effect of $P$ on the room temperature (RT) conductivity of the 2DES at the interface between {\LAO} and {\STO} is shown in Fig.~\ref{FigRevRep}. Pressure produces a smooth decrease in the sheet resistance of the 2DES, which amounts to $\sim$30{\%} for the maximum 2.44\:GPa achieved in the diamond anvil cell setup. The pressure-induced monotonous and smooth resistance decrease at room temperature ($\sim$14{\%} at 1\:GPa) was shown by all samples ($\sim$10). As shown in Fig.~\ref{FigRevRep}, the change of resistance with pressure is a reversible process. Also, cycling to 4 K under pressure was not found to cause irreversibilities. The observed reversibility and reproducibility indicate that we measure the intrinsic response of the 2DES to $P$. The decrease in resistance upon increasing $P$ at RT hints at a decrease of the electron-phonon scattering caused by pressure-induced phonon hardening. To unravel the effects of pressure on the intrinsic electronic properties of the interface we measured the effects of hydrostatic pressure on the 2DES at low temperatures. For clarity, we focus in the following discussion on the behavior at the lowest temperature used, 4~K, and examine the room-temperature behavior in the Supplement (see Fig.~\ref{Figure300K}).

\begin{figure}
\includegraphics[width=5.8cm]{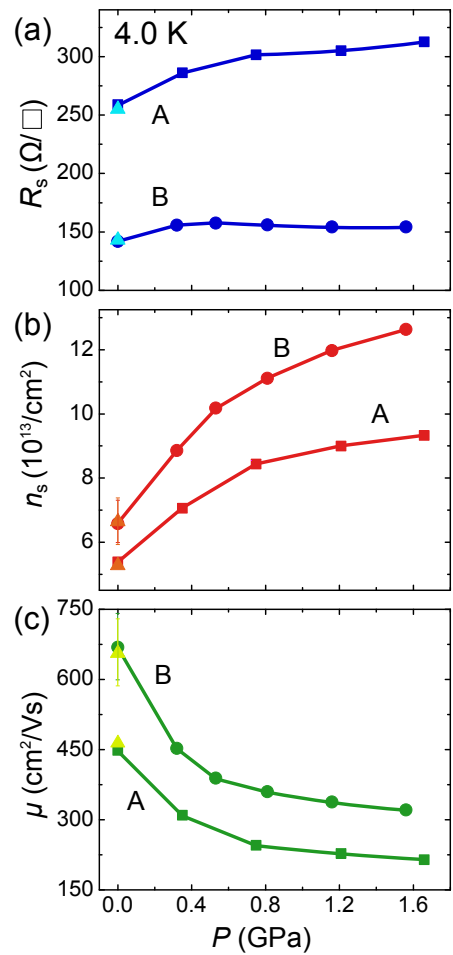}
\caption{Measured transport properties of the {\LAO}-{\STO} 2DES as a function of hydrostatic pressure. (a)
Sheet resistance, (b) sheet carrier density and (c) carrier mobility  of two {\LAO}-{\STO} heterostructures measured at 4~K as a function of applied pressure. Sample A (5 u.c. thick {\LAO}) was measured in a van der Pauw geometry, sample B (6 u.c. thick {\LAO}) by using a Hall bridge. Triangles show data taken after unloading the cell. Lines serve to guide the eye. For clarity, only those error bars that exceed the bullet size are drawn. The corresponding properties of the samples measured at room temperature are recorded in the Supplement.}
\label{Figure4K}
\end{figure}

\begin{figure}
  \includegraphics[width = 0.4\textwidth]{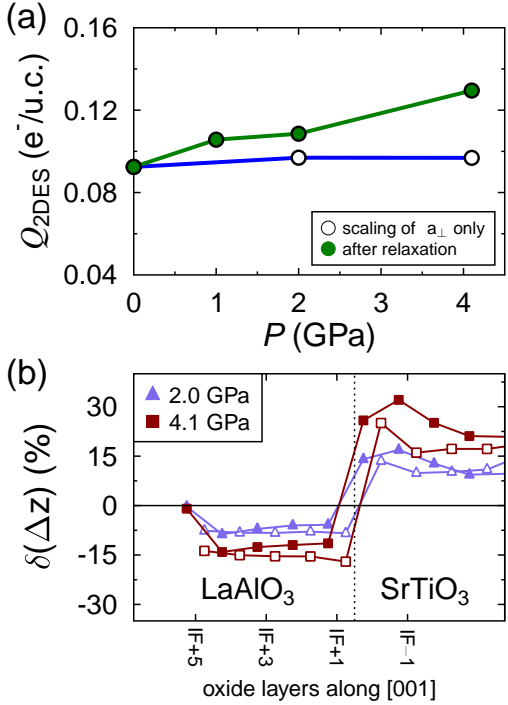}
  \caption{(a) Calculated 2DES density vs. applied hydrostatic pressure for 5 {\LAO} 
  unit cells. Open symbols represent the results of scaling-only simulations of pressure, and 
  filled symbols demonstrate the effect of the additional internal relaxation 
  under pressure. (b) Relative change of the ionic displacements 
  $ \delta (\Delta z) = (\Delta z' - \Delta z)/\Delta z \times 100\% $ in different layers 
  of the fully relaxed ({\LAO})$_5$/({\STO})$_{8.5}$/({\LAO})$_5$ heterostructure 
  (only half of the cell is shown) in response to hydrostatic pressure; open symbols 
  represent the \textit{A}O layers and filled symbols~-- the \textit{B}O$_2$ layers; 
  $\Delta z'$ stands for ionic displacements at the considered pressure ($P\neq 0$), 
  $\Delta z$ corresponds to zero pressure.}
  \label{f:pressure-effects}
\end{figure}

Whereas the sample resistance decreases with pressure at 300 K, hydrostatic pressure increases the resistance slightly at low temperatures (Fig.\ref{Figure4K}(a)). Multiple samples consistently exhibit this increase in sheet resistance at 4 K, both in van der Pauw and Hall bar configurations. The extent of the increase of resistance at 4 K, however, can vary, presumably depending on the quality of the sample (see the Supplement Fig.~\ref{FigSampleC} and Fig.~\ref{Fig_RvsT_AB} for more details). The increase of the sample resistance goes hand in hand with a reversible doubling of the sheet carrier density $n_{s}$ (Fig.~\ref{Figure4K}(b)). As a side note it is worth mentioning that magnetic field sweeps beyond  $\pm$3~T reveal in this magnetic field range an obvious multiband behavior consistent with the high carrier density shown by these samples and in agreement with other studies.~\cite{Joshua2012} At smaller fields the characteristics are pronouncedly more linear (Fig.~\ref{FigHallAB} in the Supplement). The trend observed under $P$ is nevertheless captured by the simplified one-band analysis shown in this work.\footnote{A more detailed analysis of the multiband character evolution with $P$ will be shown elsewhere (P. Seiler, J. Zabaleta \textit{et al.} in preparation).} The measured increase of $n_{s}$ is compensated by a large pressure-induced reduction of the mobility $\mu$ (Fig.~\ref{Figure4K}(c)). Interestingly, a comparable relation between $n_{s}$ and $\mu$ was revealed in a study by Y. Xie and coworkers.~\cite{Xie2013} In that work, modifications of the {\LAO} surface led to an inverse proportionality between $n_{s}$ and $\mu$. Recent pressure experiments by D. Fuchs \textit{et al.} in {\LAO}-{\STO} bilayers provide evidence of a similar response of the carrier density and mobility to hydrostatic pressure.~\cite{Fuchs2015} Note, however, that the sample from Ref.~\onlinecite{Fuchs2015}, grown at lower temperatures and oxygen partial pressures, is intrinsically different to the ones investigated here. With a {\LAO} layer five times thinner, our samples are twice as conducting and have a larger carrier density at ambient pressure. 

In order to clarify the mechanism behind the drastic changes of the 2DES observed in our experiment,
we proceed in what follows with our {\it ab initio} DFT-based theoretical investigation.


\subsection{Results of the simulations} \label{ResSim}
\begin{figure}
  \includegraphics[width = 0.4\textwidth]{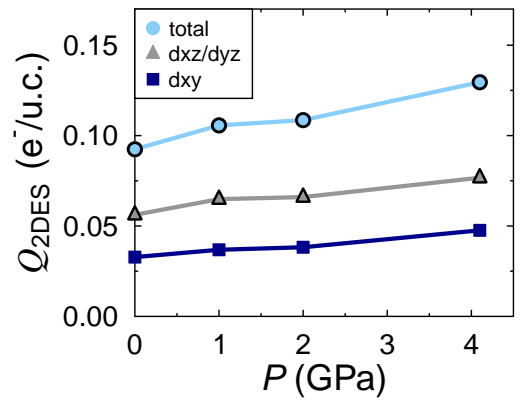}
  \caption{Orbital-resolved 2DES charge density vs. applied pressure. 
  Similar to Fig.~\ref{f:pressure-effects}(a), the effect of the additional lattice 
  reconstruction under pressure is included here. The charge density represents the sum over all {\STO} layers.}
  \label{f:density-orbitals}
\end{figure}

Starting from the fully relaxed structure with $a_\perp =
3.891\:\mathrm{\AA}$, which corresponds, in our calculations, to zero
external pressure,\footnote{Note that this pressure zero may be shifted
  with respect to the experiment.} we investigate the effects of
hydrostatic pressure by scaling down  the lattice vectors 
 by a certain percentage $ \delta = \Delta a_\perp / a_\perp
\times 100\% $ and we first keep fixed the internal atomic positions.  
Values of $ \delta $ equal to 0.5\% and 1.0\% 
correspond to the application of an isotropic pressure of $P=2.0$~GPa
and $P=4.1$~GPa, respectively, as determined from the \textit{ab initio} 
stress tensor. This simple procedure  does not reveal any 
changes in the 2DES distribution and density as a function of pressure, 
as depicted in Fig.~\ref{f:pressure-effects}(a) (empty symbols). We then
perform a further relaxation of internal coordinates (keeping fixed 
the AlO$_2$ surface layers  positions~\footnote{Fixing the two surfaces is needed, since otherwise
  the full relaxation of all internal positions would have released
  the pressure in the growth direction due to the existence of the
  vacuum side.}) in order to describe
  possible lattice reconstructions under applied pressure.
   Our results indicate that this additional relaxation
 has a large impact on the 2DES density (Fig.~\ref{f:pressure-effects}(a)
 (full symbols)). First of all,
it leads to sizable changes of the ionic displacements in both
materials as presented in
Fig.~\ref{f:pressure-effects}(b) where the relative change
of the ions displacements 
 $ \delta (\Delta z) = (\Delta z' - \Delta z)/\Delta z \times 100\% $,
($\Delta z'$ stands for ionic displacements at the considered pressure ($P\neq 0$),
and  $\Delta z$ corresponds to zero pressure) is shown for two
representative pressure values of 2.0 GPa and 4.1 GPa.
  Lattice distortions in {\LAO}
are reduced by 15\% for the largest studied pressure, whereas they
increase in {\STO} by more than 25\% (a sketch of this effect is also shown 
in Fig.~\ref{FigLAOSTO}). It should be mentioned here that
the ionic displacements in {\STO} are much smaller than those in {\LAO} (see Fig.~\ref{f:lat-reconstr}(b)). The
observed structural changes are accompanied by an enhancement of the
integrated interface charge, localized in the {\STO} portion, from
0.09~$e^-$/u.c. for 0~GPa to 0.13~$e^-$/u.c. for 4.1~GPa,\footnote{Additional
  relaxation with fixed positions of the {\LAO} surfaces changes the
  stress tensor, which becomes slightly anisotropic. However, the
  degree of anisotropy is so small that one can still speak of 
  isotropic hydrostatic-like pressure conditions.} i.e. by
approximately 45\% (see Fig.~\ref{f:pressure-effects}(a)). The
charge increases almost linearly as a function of the applied
pressure, although there is apparently a plateau near 1~GPa which 
compares well with the saturation region for the experimental curve on Fig.~\ref{Figure4K}. 
The characteristic charge distribution, shown in Figure~\ref{f:lat-reconstr}(d) 
for the zero-pressure case, does not change qualitatively for larger pressures. 
The n-type charge carriers in this slab are distributed over the whole {\STO} film, 
and p-type charges with the same total density are localized within a few atomic
layers near both {\LAO} free surfaces. For each value of pressure, 85\% of the 
local 2DES charge on the interfacial Ti cations is accommodated
by the $d_{xy}$ orbitals. The situation is opposite far from the
interface, where the $d_{xz}/d_{yz}$ states dominate. Summed over all
{\STO} layers, the total charge of the 2DES contains around 35\% $d_{xy}$
character and the rest has $d_{xz}/d_{yz}$ character. 
Under pressure, the charge densities of both the $d_{xy}$ interface states and 
the $d_{xz}/d_{yz}$ states (in-depth propagating into the bulk part of {\STO}) 
increase (Fig.~\ref{f:density-orbitals}), giving rise to the 
aforementioned overall enhancement of the 2DES carrier concentration by 45\%. The layer- and orbital-resolved charge profiles for the 
thicker slab with 20.5 STO u.c. at zero pressure and $P=2.1\:\mathrm{GPa}$ are shown 
in Fig.~\ref{f:charge-20p5}.

\begin{figure}
  \centering
  \includegraphics[width = 0.49\textwidth]{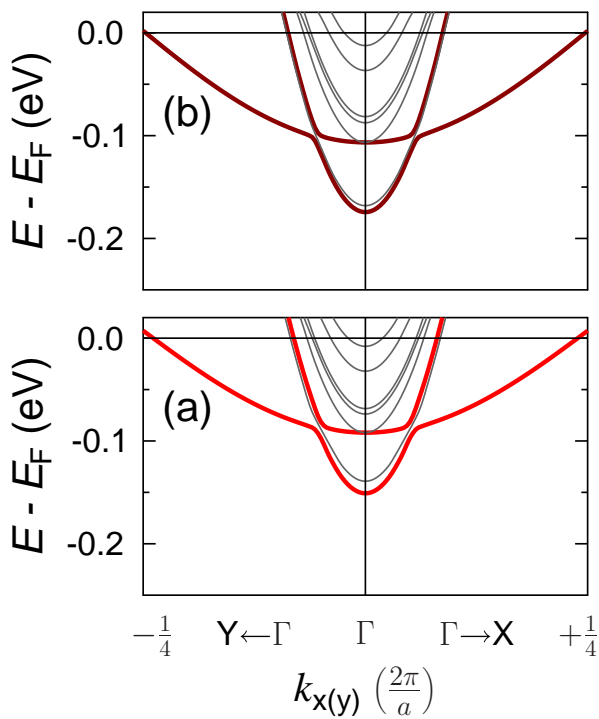}
 \caption{Band structure of the n-type {\LAO}-{\STO} heterostructure 
    for the (a) zero-pressure case and (b) $P=$~4.1~GPa. The lowest 
    occupied Ti \textit{d} conduction bands are indicated by the thick lines, 
    whereas the background band structure is plotted using thin gray lines. 
    The lowest light band has the largest weight on the interfacial Ti $d_{xy}$ states, while the
    single heavy band has contributions from all TiO$_2$ oxide layers and
    is composed mainly of the $d_{xz}/d_{yz}$ states of the more distant Ti sites.
    The site-resolved analysis of bands was performed \-using the \textit{PyPro\-car}
    code.\cite{PyProcar}}
  \label{f:band_structure}
\end{figure}

Analyzing the curvature of the light $d_{xy}$  and heavy $d_{xz}/d_{yz}$
bands near the $\Gamma$ point of the Brillouin zone (Fig.~\ref{f:band_structure}) 
we determine the effective electron masses in units 
of the electron rest mass ($m^*/m_e$). At zero pressure, $m^*/m_e$ for the 
light and heavy bands are 0.46 and 4.17 while at 4.1~GPa, the values are 
0.44 and 3.88 respectively. These results would suggest a  slight enhancement 
of the mobility of the 2DES charge carriers, which contradicts, however, 
the low-temperature experimental observations in the present work. 
In the following, we will try to resolve this puzzle by analyzing 
the polarization in {\LAO} as well as the dielectric response in {\LAO} and {\STO}
as a function of pressure.

\subsubsection{Polarization in \LAO}
We performed  a Berry phase analysis~\cite{KingSmith1993} of the electronic polarization
of selected structures
which, together with the ionic contribution, describe the response of
{\LAO} to the interface polarity and strain conditions.

\begin{figure}
  \includegraphics[width = 0.49\textwidth]{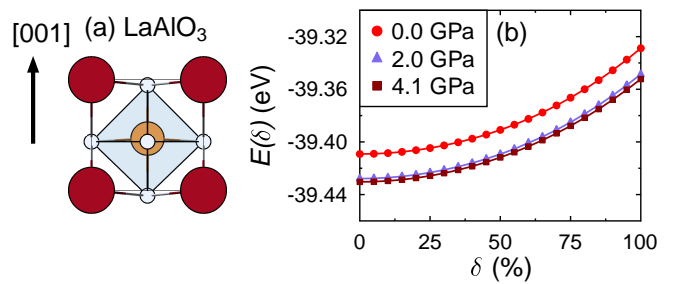}
  \caption{ (a) Unit cell of the distorted {\LAO} and (b) the potential
    profile $ E(\delta) $  for the transition between the
    centrosymmetric ($\delta = 0\:\%$) und polarized ($\delta =
    100\:\%$) tetragonal phases.}
  \label{f:LAO-cell}
\end{figure}

In order to estimate the polarity within the relaxed {\LAO}-{\STO}
heterostructure, we cut the third unit cell of {\LAO} away from the
interface (IF+3) out of the slab and form an artificial new bulk
material (Fig.~\ref{f:LAO-cell}) with the distortion of the
heterostructure. Important features, captured by this artificial bulk
{\LAO}, are the tetragonal distortion along the [001] direction as well
as the La-O and Al-O ionic displacements. The latter are zero in the
normal bulk {\LAO} and can be stabilized only in the heterostructure. 
These structural features are similar for all {\LAO} unit cells along 
the growth direction, so we consider the newly constructed artificial 
bulk as representative of the whole {\LAO} film.

Both in LaO and AlO$_2$ atomic planes, the displacements between
cations and oxygen atoms are positive along the [001] axis 
(Fig.~\ref{f:LAO-cell}) oriented away from the interface. In order to
calculate the dipole moment of this system, we define a
centrosymmetric counterpart with the same lattice vectors where all
displacements are zero. Starting from this high-symmetry phase, one
can gradually approach the low-symmetry phase through linear
interpolation of atomic coordinates (see, for example, 
Ref.~\onlinecite{Spaldin2012}). The energy profile calculated along
the deformation path (Fig.~\ref{f:LAO-cell}(b)) indicates that the
low-symmetry phase is unstable in the bulk since the high-symmetry
phase is 80~meV/u.c. lower in energy. However, for each intermediate
structure, one can estimate the electric dipole moment and, therefore,
determine the total change of polarization along the aforementioned
path. The latter value is taken as the polarization of the distorted 
{\LAO} structure.
\begin{table}
  \caption{Total lattice polarization $p$ of {\LAO} (in units of 
    $\mathrm{\mu C/cm^2}$ and e\,$\cdot$\,\AA{}) and the interface 
    charge density $\sigma$ calculated as functions of the hydrostatic pressure 
    applied on the {\LAO}-{\STO} system with 5 u.c. of \LAO}
\centering
\setlength{\tabcolsep}{10pt}
\renewcommand{\arraystretch}{1.5}
   \begin{tabular}{cccc}\hline\hline
     $P$ (GPa) & $p$ ($\mathrm{\mu C/cm^2}$) & $p$ (e\,$\cdot$\,\AA) & $\sigma$ ($e^-$/u.c.) \\ \hline
         0     &    34.5   &  1.22  &  0.09  \\
        2.0    &    31.5   &  1.12  &  0.11  \\
        4.1    &    30.1   &  1.04  &  0.13  \\ \hline\hline
   \end{tabular}
   \label{t:polarization}
\end{table}

Using this procedure, we find that both ionic and electronic contributions 
are non-vanishing in the studied case. For the {\LAO} cell taken from the fully
relaxed heterostructure with $n=5$ {\LAO} layers and at two pressures
(2.0 and 4.1~GPa), the total values of the lattice polarization are
presented in Table~\ref{t:polarization}, compared to the zero-pressure
case. The observed decrease of the lattice polarization for larger
pressures is consistent with a considerable suppression of lattice
distortions (Fig.~\ref{f:pressure-effects}(b)). In general, the
polarization is oriented towards the free {\LAO} surface, which means
that it partially screens the interface charge. This explains why the
observed densities of the 2DES at $P = 0$ are almost an order of
magnitude smaller than the value of 0.5~$e^-$/u.c predicted by the
polar catastrophe model. Understandably, the 2DES density becomes
larger when the polarization of {\LAO} and, therefore, the lattice
screening are reduced under pressure. If the lattice distortions were
not present, then the screening would be close to zero and the
interface charge $Q$ would approach its maximal value of 0.5~$e^-$, as
expected from the polar catastrophe scenario in its simplest version and confirmed by the large 
carrier density obtained in our calculations for the unrelaxed heterostructures
(Fig.~\ref{f:lat-reconstr}(c)). In reality, however, the finite thickness of 
{\LAO} together with structural relaxation effects lead to a non-vanishing induced 
lattice polarization. We expect our \textit{ab initio} data for this polarization to better agree  
 with our measurements, which were done in the absence of photo-illumination, 
since the latter induces additional carriers that screen the polar discontinuity
and suppress the {\LAO} polarization as well as the internal field, as
found in Refs.~\onlinecite{Berner2013a} and \onlinecite{Segal2009}.

In view of the above discussion, one can reconsider the effect of the
GGA+$U$ corrections on the lattice reconstruction and electronic properties 
of the 2DES. In particular, we find a substantial increase by more than 200\% 
of the carrier density as the $U_{\rm eff}$ parameter on the La $4f$ states 
varies between 0 and 8~eV (results not shown here). The origin of this strong 
dependence can be found in the structural features. Intralayer displacements 
in {\LAO} are moderately suppressed for $U_{\rm eff}(\text{La}) = 8\:\text{eV}$, 
which, according to the previous discussion, effectively reduces the screening 
of the polar discontinuity and enhances the charge density at the interface.

\subsubsection{Dielectric properties}
The dielectric response, both
from the lattice and from the electronic subsystems, was obtained
using the formalism described in Refs.~\onlinecite{Delugas2005} and~\onlinecite{Gajdos2006}.

Whereas the behavior of the lattice polarization in {\LAO} under pressure
explains the experimental trends of the 2DES density, another mystery
remains to be solved. Experimentally, we find that the interface resistivity increases as a function of pressure, although the carrier density becomes larger. This implies that the mobility of the electron system 
is significantly reduced. However, at the beginning of this section we have seen that 
the observed changes in the band structure under pressure cannot explain 
this reduction. In searching for an explanation to this puzzle, we turn our attention to the pressure-dependent dielectric permittivity of the {\LAO} and {\STO} portions of the studied heterostructure, which are arguably different from their bulk counterparts. The dielectric properties have an impact on the screening of charged impurities and therefore affect the sample conductivity. 

In the first step, we study the dielectric response of the distorted
``bulk'' \LAO,  in view of the polarization effects discussed
above.  One has to distinguish between the contribution of the lattice 
($\varepsilon_\mathrm{ion}(\omega)$) to the complex
permittivity $\varepsilon(\omega)$, which prevails at low frequencies,
and the electronic dielectric constant
$\varepsilon_\mathrm{el}(\omega)$, which shows oscillations at higher
frequencies, usually in the THz region.  Our results indicate that the
electronic tensor $\varepsilon_\mathrm{el}(\omega)$ is diagonal with
slightly different in-plane and out-of-plane components due to the
tetragonal distortions in the \LAO. The effect of increasing pressure
is almost negligible for $\varepsilon_\mathrm{el}(\omega)$ (not shown
here): characteristic peaks shift only slightly towards higher
frequencies, and the static permittivity reveals almost no dependence
on pressure.

\begin{figure}
  \centering
  \includegraphics[width = 0.49\textwidth]{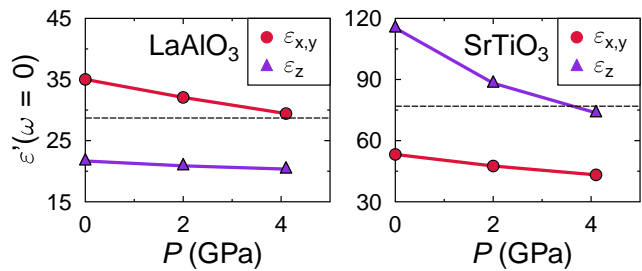}
  \caption{Pressure dependence of the static dielectric permittivity $\varepsilon'(\omega = 0)$ 
     (real part) of {\LAO} and {\STO}. In each case, the structure is taken from the 
     middle unit cell of the corresponding material in the fully relaxed (\LAO)$_5$/(\STO)$_{8.5}$/(\LAO)$_5$ 
     slab, shown in Fig.~\ref{f:lat-reconstr}(a). The values of $\varepsilon'$ for bulk oxides 
     with their native cubic structure are marked by the dashed horizontal lines. Off-diagonal elements 
     are negligible and are, therefore, not presented here. In each panel, the lines connecting 
     points on the plot are guides to the eye.}
  \label{f:dielectric-properties}
\end{figure}

A larger effect can be expected from the lattice response of {\LAO}, which is well
known to give the largest contribution to the 
static dielectric permittivity.~\cite{Delugas2005}
It can be determined from the vibrational
modes and Born effective charges. Due to the epitaxial strain, the
dielectric permittivity is an anisotropic tensor, which turns out,
however, to be diagonal, similar to $\varepsilon_\mathrm{el}(\omega)$. 
The anisotropy makes the in-plane components $ \varepsilon_{x,y} $ 
larger and the out-of-plane component $ \varepsilon_z $ smaller than 
the bulk value (28.7 in our calculations for the equilibrium lattice volume).  
Our findings reveal a large pressure-induced reduction of all components 
of the ionic dielectric tensor, in contrast to the electronic
counterpart, which shows almost no changes. At the end, this gives a sizable suppression of the
total static permittivity tensor with applied pressure (see
Fig.~\ref{f:dielectric-properties}). The in-plane components $
\varepsilon_{x,y} $ are most affected in this process, and the
out-of-plane component $ \varepsilon_z $ remains practically constant,
i.e. the {\LAO} oxide becomes dielectrically more isotropic with growing
pressure, which must be due to the suppressed ionic displacements.

We perform an analogous study for the {\STO} substrate. A crucial
difference, however, is the electronic structure, which is metallic
throughout the whole {\STO} layer in the case of 8.5 {\STO}
unit cells in the slab. For that reason, if we consider a
local cut of its structure, its bulk insulating properties cannot be directly
related to those of the heterostructure. Nevertheless, we might
consider, in the first approximation, the effect of lattice
deformation in {\STO} under pressure on the dielectric tensor, calculated
for the insulating bulk phase with a structure taken from the central
unit cell of the slab. Due to the mirror symmetry, the intralayer
$\Delta z$ are zero, so that only tetragonal distortions are
present. In Fig.~\ref{f:dielectric-properties} (right panel), the calculated
out-of-plane component $ \varepsilon_z $ drops rapidly against
external pressure, while the in-plane component shows much smaller
changes. This is the opposite situation compared to {\LAO}, where the
in-plane $ \varepsilon_{x,y} $ showed the largest response to
pressure. However, both materials demonstrate a substantial
suppression of their dielectric constants as the hydrostatic pressure
is applied.

\begin{figure}
\includegraphics[width=6.8cm]{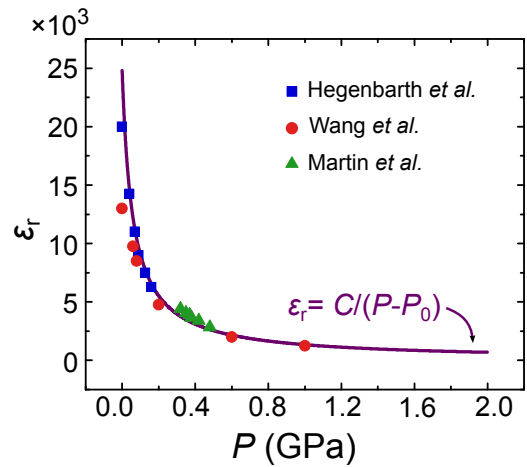}
\caption{Measured pressure dependence of low-temperature dc-dielectric constant of {\STO} single crystals as compiled from the literature.\cite{Hegenbarth1967, Martin1975, Wang2000a} The data of reference \cite{Martin1975} were measured at a bias field of 2.5 kV/cm, the other data without a bias field. The original data were taken at temperatures close to 4.2 K, but not necessarily at 4.2 K. If necessary we interpolated or extrapolated the measured data to 4.2 K. The solid line presents a fit to the function $\epsilon_{r}$($P$)=$C$/($P$-$P_{0}$) with $C$=1420\;GPa\;$\pm$\;60\;GPa and $P_{0}$=-0.06\;GPa\;$\pm$0.01\;GPa.}
\label{f:STO_eps}
\end{figure}

We note that our calculated structures show an almost
constant $c/a$ ratio for the studied pressure range, which means that
the {\STO} lattice parameter is simply scaled down with increasing
pressure. In this respect, we can compare our results with related
experiments\cite{Samara1966} on bulk {\STO} with cubic symmetry, 
summarized in Fig.~\ref{f:STO_eps}. These experiments demonstrated 
a pressure-dependent dielectric constant $\varepsilon' (P) = C / (P - P_0)$ 
that decreases monotonically with pressure, similar to a Curie-Weiss law.

At this point, we can address the issue of reduced mobility of the
2DES.  As we discussed previously, due to the suppression 
of lattice polarization in {\LAO} under pressure the screening of the polar 
discontinuity is reduced. As a result, the density of the interface
charge carriers is enhanced. If crystal defects or impurities were present in both 
parts of the heterostructure, they would serve as scattering centers 
for the charge carriers from the 2DES. By increasing pressure, we find 
a clear decrease of the dielectric constants of both {\LAO} and {\STO}. 
This results in a less effective screening of crystal defects by the 
lattice ions and, consequently, a higher scattering rate for the mobile 
charge carriers. We argue that this might be the origin of the reduced mobility 
that leads to an increase in sheet resistivity, as shown in Fig.~\ref{Figure4K}(a,c). 

\section{Concluding remarks}
In this paper, we performed a combined experimental and theoretical study 
of the hydrostatic pressure effects on the electronic properties of 
the 2DES in {\LAO}-{\STO}~(001). On the one hand, experimental results 
of this work show that the density of the electron system is considerably 
enhanced by moderate pressures, i.e. by almost 100\% at $\sim$1.6\,GPa. 
This phenomenon can be explained by the induced lattice reconstruction 
of the {\LAO} film, which acts against the interface charge and can be partially 
suppressed by the external pressure, as demonstrated by our first-principle analysis. 
Other effects by which pressure alters the carrier density cannot be discarded, 
but our measurements did not give indications of their existence. 
On the other hand, the calculated static dielectric permittivity decreases 
under the influence of pressure for both materials ({\LAO} and {\STO}), which 
at least partly accounts for the experimental observations of the enhanced 
interface resistivity.

The interface screening is an orbital-dependent phenomenon: clearly,
the $d_{xz}/d_{yz}$ states propagate much further, into deeper layers
of {\STO}, whereas the $d_{xy}$ states are more localized within 3--4
u.c. near the interface. Our calculated results with thick slabs indicate that
7~u.c. of {\STO} are involved in the formation of the 2DES. An important
issue here is the band gap of both oxides. In our GGA+$U$ calculations
with $U = 4\:\mathrm{eV}$ on the Ti \textit{d} states, the band gap of
bulk {\STO} is estimated to be equal to 2.4~eV, which is closer to the
experimental value (3.2 eV) than the standard GGA estimate (1.8~eV). 
Nevertheless, the delocalization effect might be slightly overestimated 
in our calculations. This should not affect qualitatively the statements 
concerning the total electron density changes in response to the external pressure.

Further open issues are, for instance, point defects and their role
under high pressures. Recently, polarity-induced defect formation at
ambient pressure was discussed as an alternative origin of the 
2DES\cite{Yu2014,Zhou2015A} in order to resolve certain purported shortcomings
of the polar catastrophe scenario. It cannot be completely ruled out that, under pressure, these defects 
 produce additional carriers . We believe, however, that our calculations of the ideal {\LAO}-{\STO} structure still 
uncover relevant processes that occur under pressure.  Whereas the defect-free 
{\LAO} and {\STO} layers in our study demonstrate large dielectric changes 
under pressure, which affect both the density and the mobility of the 2DES, 
it might well happen that the concentration and distribution of defects in 
both materials are not constant\footnote{Experimental results of our work and 
 Ref.~\onlinecite{Fuchs2015} demonstrate, however, that the observed 
 pressure effects are reversible.} as well, which would have an important 
effect on the dielectric response. The observed increase of the 2DES 
density would be further enhanced when the negatively charged oxygen defects, 
possibly present at the AlO$_2$ surface,~\cite{Berner2013a,Berner2013b} 
multiply at higher pressures. Verification of this hypothesis merits 
further study.

\section*{ACKNOWLEDGEMENTS}
Part of this work was financially supported by DFG Sonderforschungsbereich TRR 49 and TRR 80, 
and Research Unit FOR 1346. We would like to thank M. Altmeyer, H. Boschker, M. Lanzano,
and U. Zschieschang for many useful discussions. The authors thank U. Engelhardt, A. G\"uth, 
I. Hagel, M. Hagel, C. Hughes, Y. Link, M. Schmid, and B. Stuhlhofer for their support, 
and B. Keimer and R. Kremer for access to the piston cylinder cell and the physical property 
measurement system. The computer time was allotted by the centre for supercomputing (CSC) 
in Frankfurt and by the computer center of Goethe University.

\onecolumngrid

\FloatBarrier
\clearpage\newpage

\begin{center}
\textbf{\large{Hydrostatic pressure response of an oxide two-dimensional electron system}}
\vspace{12pt}

 J.~Zabaleta,$^1$ V.S.~Borisov,$^2$ R.~Wanke,$^1$ H.O.~Jeschke,$^2$ S.C.~Parks,$^1$ B.~Baum,$^1$ A.~Teker,$^1$\\[2pt] T.~Harada,$^1$ K~Syassen,$^1$ T.~Kopp,$^3$, N.~Pavlenko,$^3$ R.~Valent\'i,$^2$ and J.~Mannhart$^1$
\vspace{4pt}

\small{
\textit{
 $^1$Max Planck Institute for Solid State Research, 70569 Stuttgart, Germany\\[1pt]
 $^2$Institute of Theoretical Physics, Goethe University, 60438 Frankfurt am Main, Germany\\[1pt]
 $^3$Center for Electronic Correlations and Magnetism,\\ University of Augsburg, 86135 Augsburg, Germany\\[2pt]
 }
 }
 (Dated: \today)
\end{center}

\section*{{\large Supplementary Material}}
In the following section we compile figures of measurements and calculations as supporting information to our work.
\begin{figure}
\includegraphics[width=8.8cm]{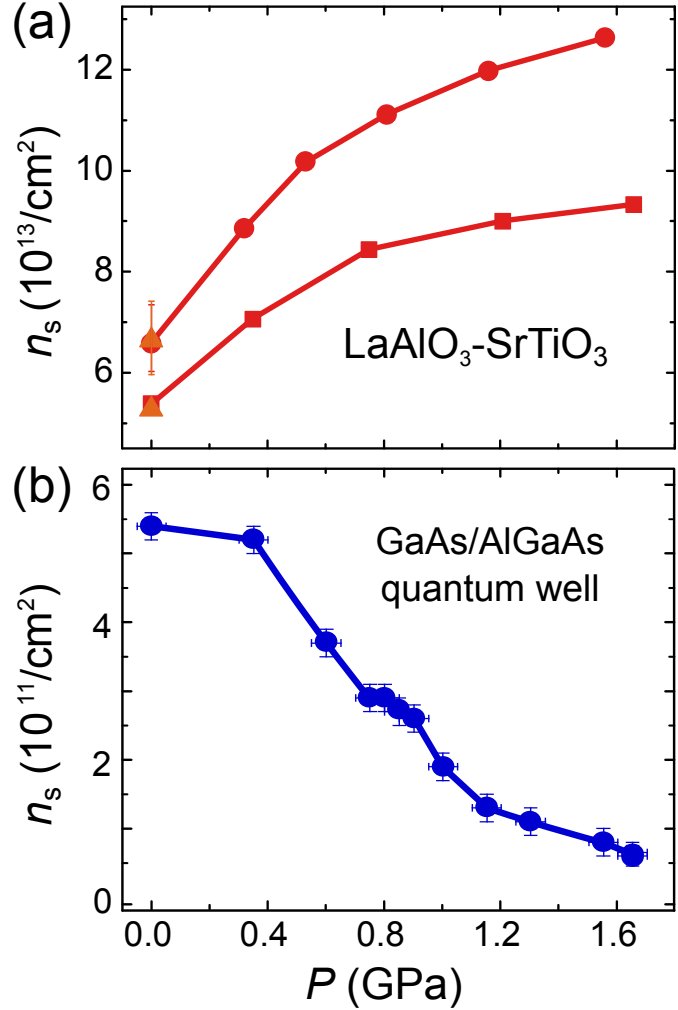}
\caption{Effect of pressure on 2D carrier densities: (a) Behavior of the \LAO-\STO interface compared to (b) a GaAs/AlGaAs single quantum well (data redrawn from Ref.~\onlinecite{Ernst1994}). Only error bars that exceed the bullet size are drawn.}
\label{Fig_AlGaAs}
\end{figure}

\begin{figure}
\includegraphics[width=9.8cm]{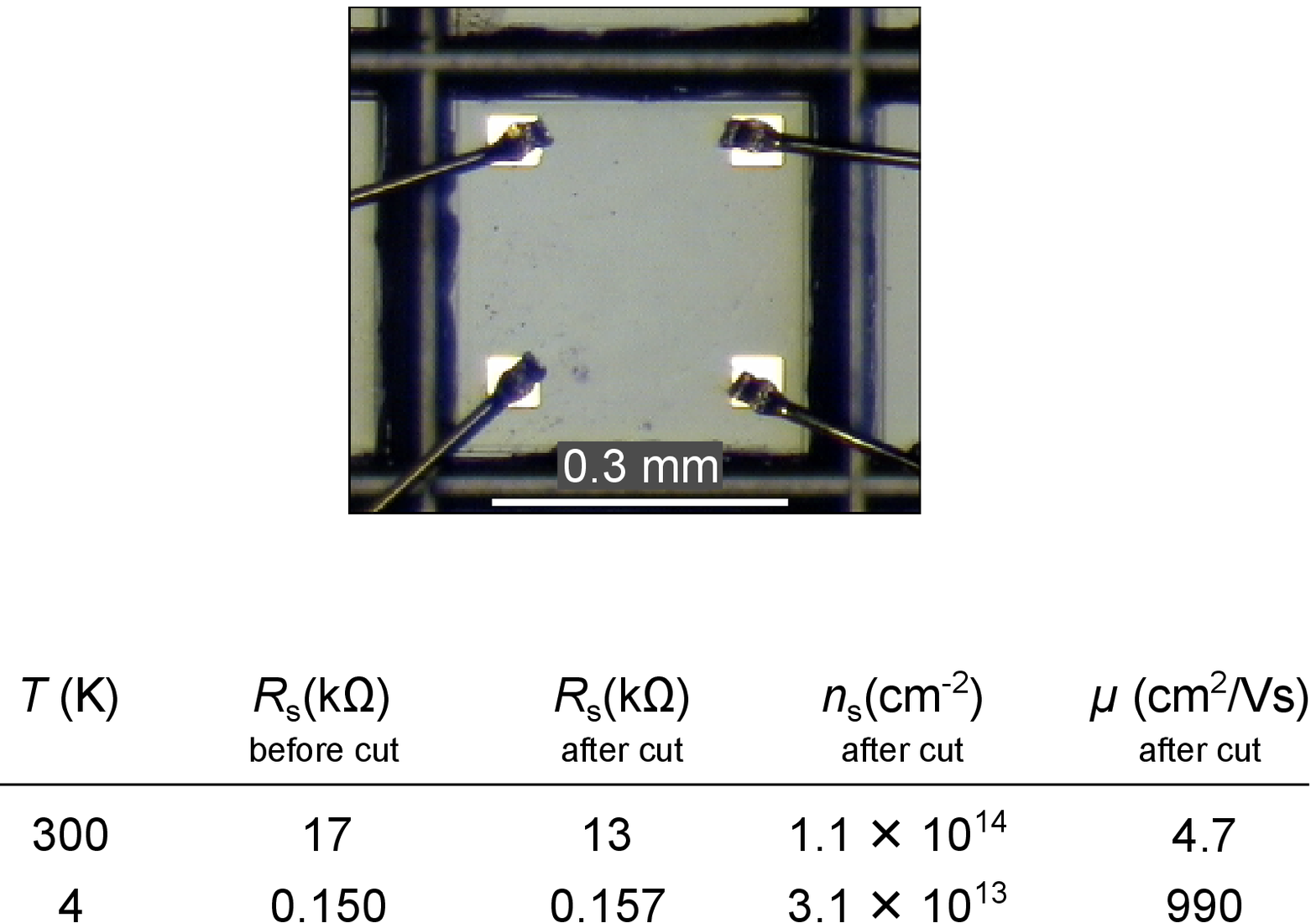}
\caption{Optical microscopy image of a cut sample wire bonded for electrical characterization, and table of the electrical characterization results of the assembly at ambient pressure.}
\label{FigBondedSample}
\end{figure}

\begin{figure}
  \includegraphics[height = 0.55\textheight]{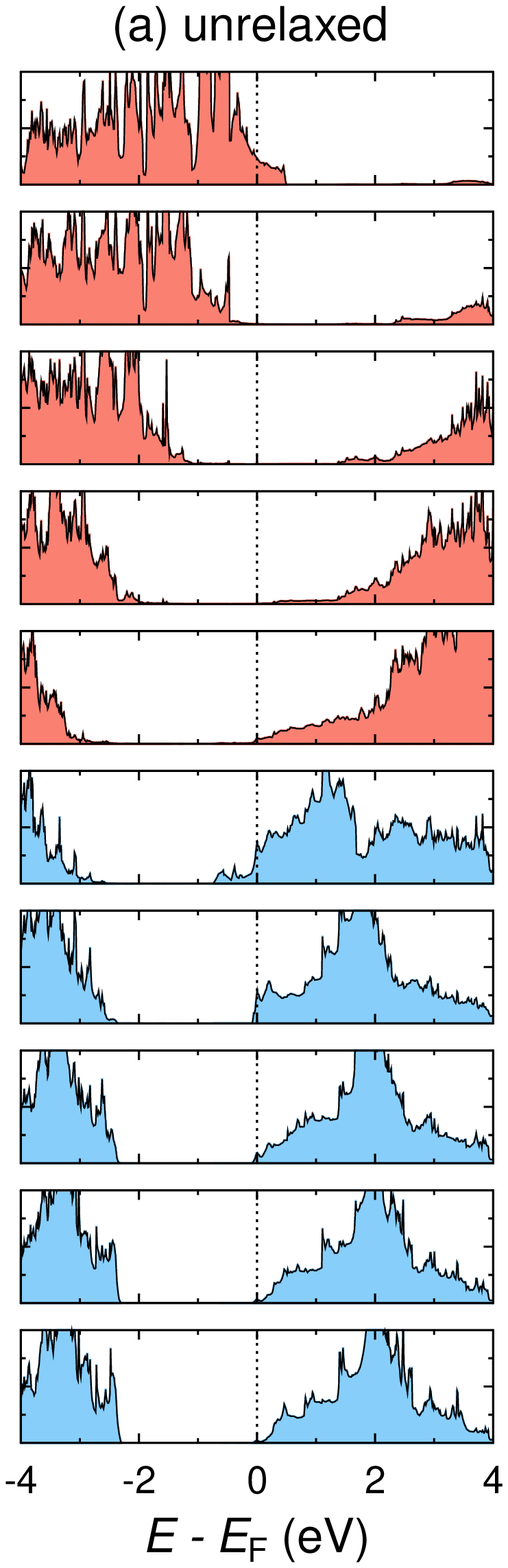}
  \includegraphics[height = 0.55\textheight]{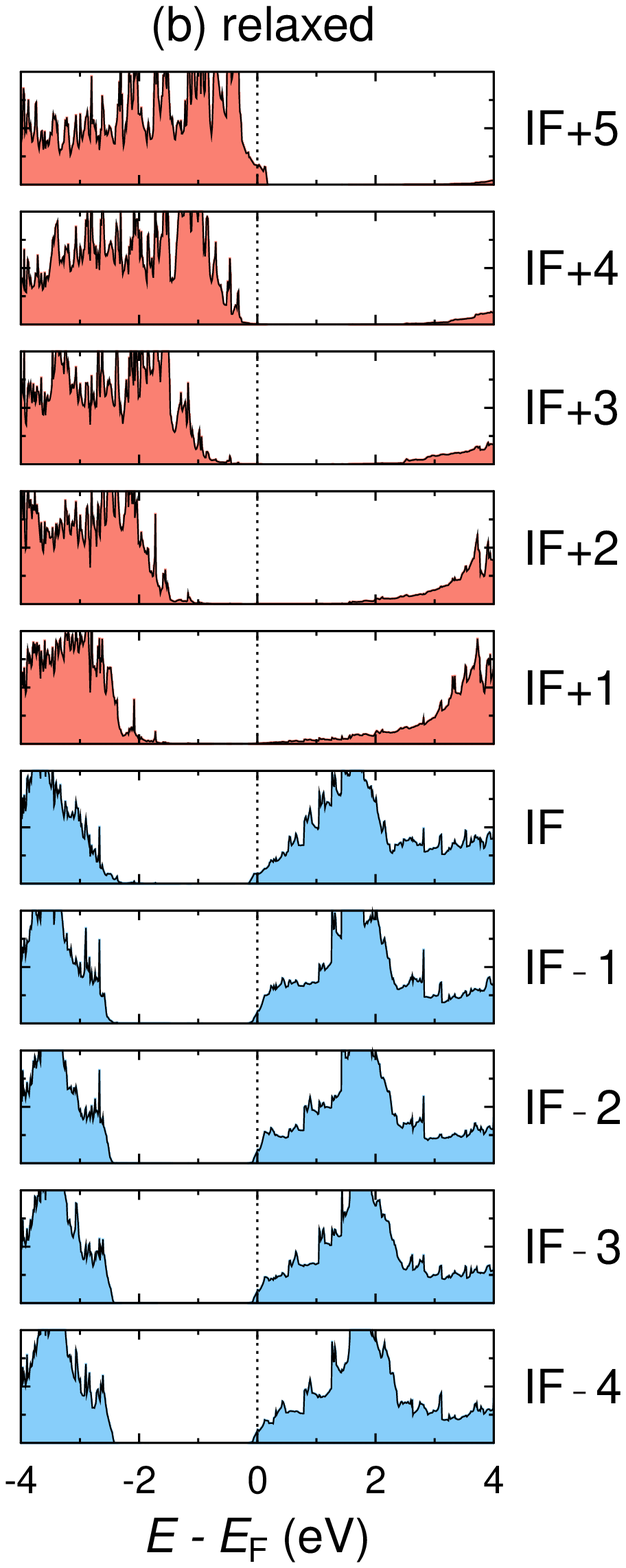}
  \caption{Unit-cell resolved density of states of the
    ({\LAO})$_5$/({\STO})$_{8.5}$/({\LAO})$_5$ slab for the unrelaxed (left
    panel) and structures relaxed internally while keeping the in-plane lattice 
    parameter fixed at $ a_\perp = 3.905\:\mathrm{\AA} $ (right panel). 
    Layers are labeled in the same way as in Fig.~3(a) from the main text.}
  \label{f:DOS-layers}
\end{figure}

\begin{figure}
  \includegraphics[width = 0.56\textwidth]{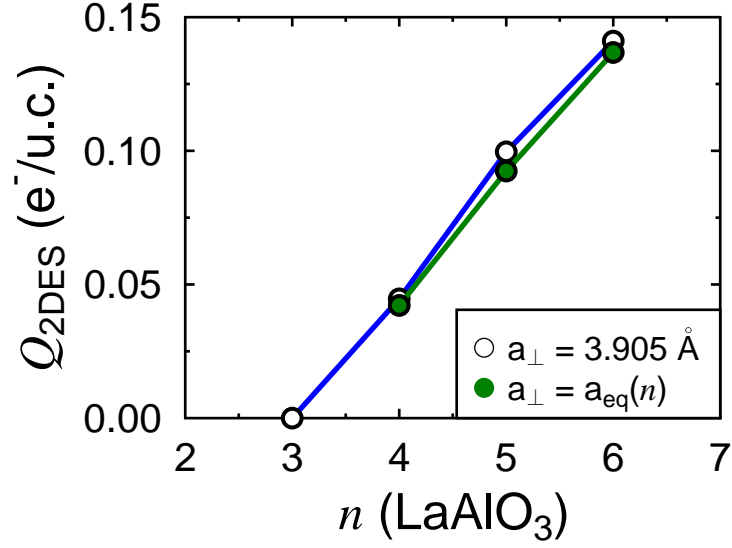}
  \caption{2DES density vs. the {\LAO} thickness. Filled symbols represent fully relaxed 
  structures with the equilibrium lattice constant $a_\mathrm{eq}$ being a function 
  of $n$ (number of {\LAO} unit cells) and open symbols stand for internally relaxed 
  structures with fixed $a_\perp = 3.905\:\mathrm{\AA}$. The lines connecting points 
  on the plot are guides to the eye.}
  \label{f:2DES-density-thickness}
\end{figure}

\begin{figure}
  \includegraphics[width = 0.56\textwidth]{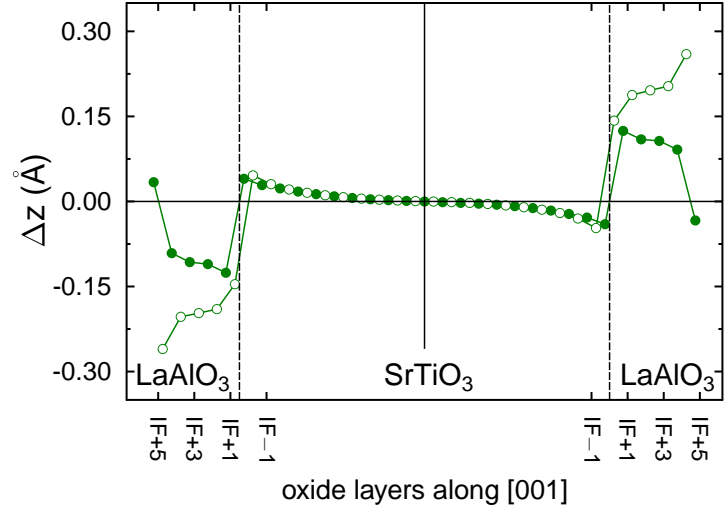}
  \caption{Ionic displacements in \textit{A}O ($A={\rm Sr}$, La; open symbols) and 
  \textit{B}O$_2$ ($B={\rm Ti}$, Al; filled symbols) oxide layers of the 
  fully relaxed (\LAO)$_5$/(\STO)$_{20.5}$/(\LAO)$_5$ slab at zero pressure.}
  \label{f:Delta-z-20p5}
\end{figure}

\begin{figure}
\includegraphics[width=8.5cm]{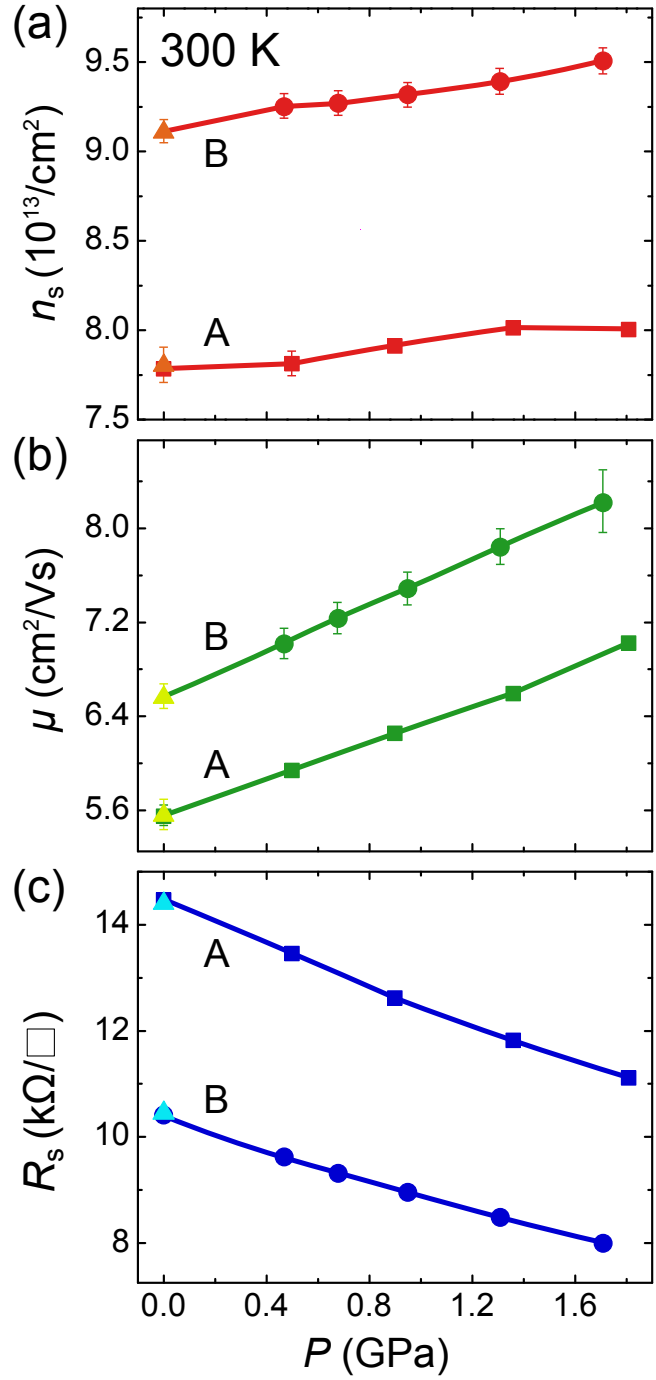}
\caption{Transport properties of the {\LAO}-{\STO} 2DES as a function of hydrostatic pressure measured at room temperature. Sheet carrier density (a), mobility (b), and sheet resistance (c) of van der Pauw and Hall bar samples as a function of applied pressure. Triangles stand for measurements after unloading the cell. For clarity, only those error bars that exceed the bullet size are drawn. Lines serve as guides to the eye.}
\label{Figure300K}
\end{figure}

\begin{figure}
\includegraphics[width=7.8cm]{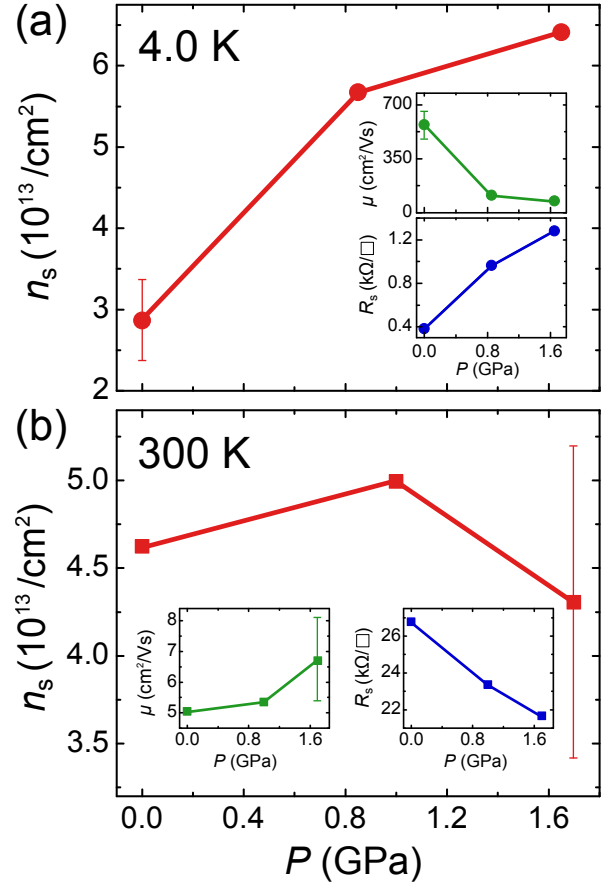}
\caption{Transport properties of a {\LAO}-{\STO} 2DES in van der Pauw configuration as a function of hydrostatic pressure. This sample showed the largest pressure-induced changes of the sheet carrier density and the carrier mobility at 4 K. The error bars reflect the differences of the results of different magnetic-field sweeps. During the 1.7 GPa warmup a contact deteriorated, resulting in the large error bar of the data point at 300 K, 1.7 GPa. Only error bars that exceed the bullet size are drawn. Lines serve as guides to the eye.}
\label{FigSampleC}
\end{figure}

\begin{figure}
\includegraphics[width=7.8cm]{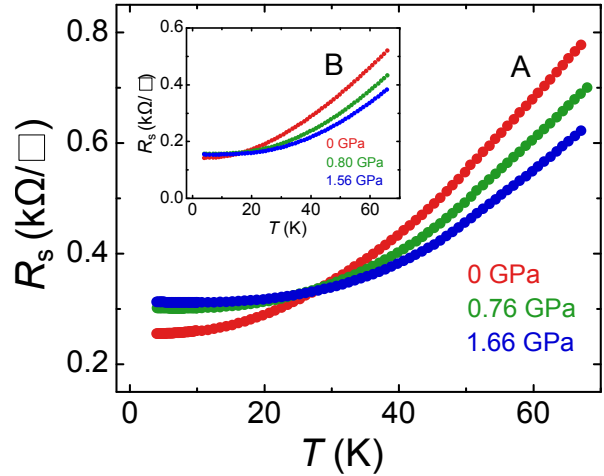}
\caption{Sheet resistance of sample A (van der Pauw configuration) and B (Hall bar configuration) measured as a function of temperature for several values of hydrostatic pressure. The two types of samples show similar characteristics.}
\label{Fig_RvsT_AB}
\end{figure}

\begin{figure}
\includegraphics[width=7.8cm]{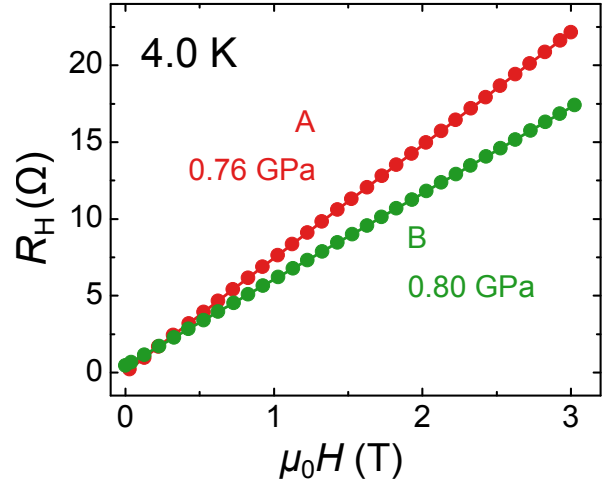}
\caption{Hall resistance as a function of the magnetic field measured on two samples (A: 5-unit-cell-thick {\LAO}, van der Pauw; B: 6-unit-cell-thick {\LAO}, Hall bridge) at 4 K under hydrostatic pressure. In order to exclude longitudinal components in the Hall resistance, the values shown here are anti-symmetrized, e.g. we plot the averaged difference between the Hall resistance measured at identical positive and negative magnetic fields.}
\label{FigHallAB}
\end{figure}

\begin{figure}
  \includegraphics[width = 0.55\textwidth]{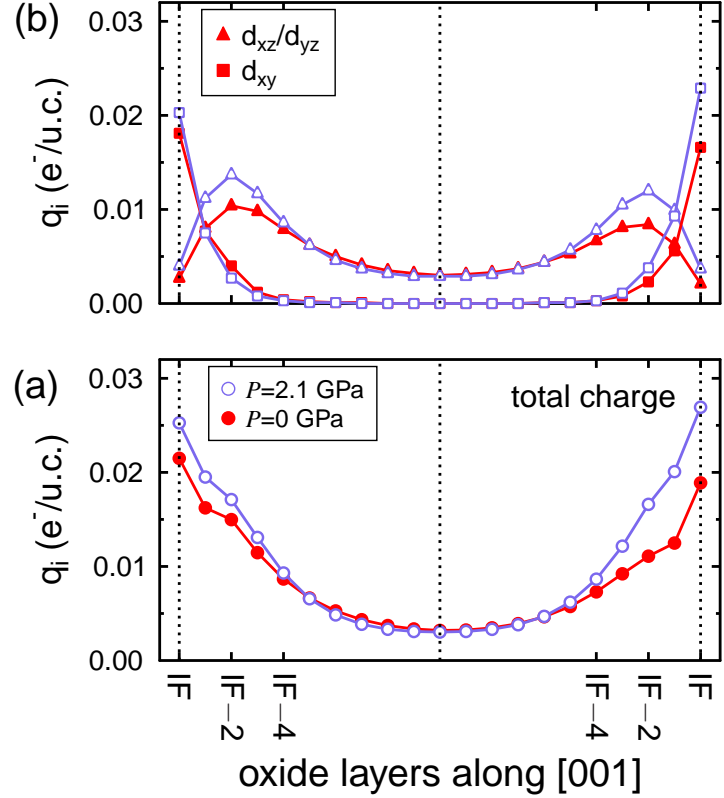}
  \caption{Charge distribution in the {\STO} part of the fully 
    relaxed (\LAO)$_5$/(\STO)$_{20.5}$/(\LAO)$_5$ slab at zero pressure and $P=2.1\:\mathrm{GPa}$. 
    Total electronic charge (a) and its $d_{xy}$ and $d_{xz}/d_{yz}$ components (b) are presented.}
  \label{f:charge-20p5}
\end{figure}

\clearpage

\bibliographystyle{prb-titles}
\bibliography{./paper}

\begin{thebibliography}{10}
\providecommand{\bibAnnoteFile}[1]{%
  \IfFileExists{#1}{\begin{quotation}\noindent\textsc{Key:} #1\\
  \textsc{Annotation:}\ \input{#1}\end{quotation}}{}}
\providecommand{\bibAnnote}[2]{%
  \begin{quotation}\noindent\textsc{Key:} #1\\
  \textsc{Annotation:}\ #2\end{quotation}}
\providecommand{\bibinfo}[2]{#2}

\bibitem{Ohtomo2004}
\bibinfo{author}{A.~Ohtomo} and \bibinfo{author}{H.~Y. Hwang},
  \bibinfo{title}{A high-mobility electron gas at the {LaAlO$_3$/SrTiO$_3$}
  heterointerface}, \bibinfo{journal}{Nature} \textbf{\bibinfo{volume}{427}},
  \bibinfo{pages}{423} (\bibinfo{year}{2004}).
\bibAnnoteFile{Ohtomo2004}

\bibitem{Paul1998}
\bibinfo{editor}{W.~Paul} and \bibinfo{editor}{T.~Suski}, editors,
  \bibinfo{title}{High pressure semiconductor physics I $\&$ II}, volume
  \bibinfo{volume}{54-55}, \bibinfo{publisher}{Elsevier Science}
  (\bibinfo{year}{1998}).
\bibAnnoteFile{Paul1998}

\bibitem{Schilling2007}
\bibinfo{author}{J.~S. Schilling}, \bibinfo{title}{Handbook of
  High-Temperature Superconductivity: Theory and Experiment}, chapter
  \bibinfo{chapter}{High-Pressure Effects}, \bibinfo{publisher}{Springer New
  York} (\bibinfo{year}{2007}).
\bibAnnoteFile{Schilling2007}

\bibitem{Drozdov2015}
\bibinfo{author}{A.~P. Drozdov}, \bibinfo{author}{M.~I. Eremets},
  \bibinfo{author}{I.~A. Troyan}, \bibinfo{author}{V.~Ksenofontov}, and
  \bibinfo{author}{S.~I. Shylin}, \bibinfo{title}{Conventional
  superconductivity at 203 kelvin at high pressures in the sulfur hydride
  system}, \bibinfo{journal}{Nature} \textbf{\bibinfo{volume}{525}},
  \bibinfo{pages}{73} (\bibinfo{year}{2015}).
\bibAnnoteFile{Drozdov2015}

\bibitem{Haeni2004}
\bibinfo{author}{J.~H. Haeni}, \bibinfo{author}{P.~Irvin},
  \bibinfo{author}{W.~Chang}, \bibinfo{author}{R.~Uecker},
  \bibinfo{author}{P.~Reiche}, \bibinfo{author}{Y.~L. Li},
  \bibinfo{author}{S.~Choudhury}, \bibinfo{author}{W.~Tian},
  \bibinfo{author}{M.~E. Hawley}, \bibinfo{author}{B.~Craigo},
  \bibinfo{author}{A.~K. Tagantsev}, \bibinfo{author}{X.~Q. Pan},
  \bibinfo{author}{S.~K. Streiffer}, \bibinfo{author}{L.~Q. Chen},
  \bibinfo{author}{S.~W. Kirchoefer}, \bibinfo{author}{J.~Levy}, and
  \bibinfo{author}{D.~G. Schlom}, \bibinfo{title}{{R}oom-temperature
  ferroelectricity in strained {SrTiO$_3$}}, \bibinfo{journal}{Nature}
  \textbf{\bibinfo{volume}{430}}, \bibinfo{pages}{758} (\bibinfo{year}{2004}).
\bibAnnoteFile{Haeni2004}

\bibitem{Bark2011}
\bibinfo{author}{C.~W. Bark}, \bibinfo{author}{D.~A. Felker},
  \bibinfo{author}{Y.~Wang}, \bibinfo{author}{Y.~Zhang}, \bibinfo{author}{H.~W.
  Jang}, \bibinfo{author}{C.~M. Folkman}, \bibinfo{author}{J.~W. Park},
  \bibinfo{author}{S.~H. Baek}, \bibinfo{author}{H.~Zhou},
  \bibinfo{author}{D.~D. Fong}, \bibinfo{author}{X.~Q. Pan},
  \bibinfo{author}{E.~Y. Tsymbal}, \bibinfo{author}{M.~S. Rzchowski}, and
  \bibinfo{author}{C.~B. Eom}, \bibinfo{title}{Tailoring a two-dimensional
  electron gas at the {LaAlO$_3$/SrTiO$_3$} (001) interface by epitaxial
  strain}, \bibinfo{journal}{Proc. Natl. Acad. Sci. U. S. A.}
  \textbf{\bibinfo{volume}{108}}, \bibinfo{pages}{4720} (\bibinfo{year}{2011}).
\bibAnnoteFile{Bark2011}

\bibitem{Nazir2014a}
\bibinfo{author}{S.~Nazir}, \bibinfo{author}{M.~Behtash}, and
  \bibinfo{author}{K.~Yang}, \bibinfo{title}{Enhancing interfacial conductivity
  and spatial charge confinement of {LaAlO$_3$/SrTiO$_3$} heterostructures via
  strain engineering}, \bibinfo{journal}{Appl. Phys. Lett.}
  \textbf{\bibinfo{volume}{105}} (\bibinfo{year}{2014}).
\bibAnnoteFile{Nazir2014a}

\bibitem{Nazir2014b}
\bibinfo{author}{S.~Nazir} and \bibinfo{author}{K.~Yang},
  \bibinfo{title}{First-Principles Characterization of the Critical Thickness
  for Forming Metallic States in Strained {LaAlO$_3$/SrTiO$_3$} (001)
  Heterostructure}, \bibinfo{journal}{ACS Appl. Mater. Inter.}
  \textbf{\bibinfo{volume}{6}}, \bibinfo{pages}{22351} (\bibinfo{year}{2014}).
\bibAnnoteFile{Nazir2014b}

\bibitem{Hicks2014}
\bibinfo{author}{C.~W. Hicks}, \bibinfo{author}{M.~E. Barber},
  \bibinfo{author}{S.~D. Edkins}, \bibinfo{author}{D.~O. Brodsky}, and
  \bibinfo{author}{A.~P. Mackenzie}, \bibinfo{title}{{P}iezoelectric-based
  apparatus for strain tuning}, \bibinfo{journal}{Rev. Sci. Instrum.}
  \textbf{\bibinfo{volume}{85}}, \bibinfo{pages}{065003}
  (\bibinfo{year}{2014}).
\bibAnnoteFile{Hicks2014}

\bibitem{Gao1994}
\bibinfo{author}{L.~Gao}, \bibinfo{author}{Y.~Y. Xue},
  \bibinfo{author}{F.~Chen}, \bibinfo{author}{Q.~Xiong}, \bibinfo{author}{R.~L.
  Meng}, \bibinfo{author}{D.~Ramirez}, \bibinfo{author}{C.~W. Chu},
  \bibinfo{author}{J.~H. Eggert}, and \bibinfo{author}{H.~K. Mao},
  \bibinfo{title}{Superconductivity up to 164 K in
  {HgBa$_2$Ca$_{m-1}$Cu$_m$O$_{2m+2+\delta}$} (m=1,2, and 3) under
  quasi-hydrostatic pressures}, \bibinfo{journal}{Phys. Rev. B}
  \textbf{\bibinfo{volume}{50}}, \bibinfo{pages}{4260} (\bibinfo{year}{1994}).
\bibAnnoteFile{Gao1994}

\bibitem{Lemarrec2002}
\bibinfo{author}{F.~Le~Marrec}, \bibinfo{author}{A.~Demuer},
  \bibinfo{author}{D.~Jaccard}, \bibinfo{author}{J.~M. Triscone},
  \bibinfo{author}{M.~K. Lee}, and \bibinfo{author}{C.~B. Eom},
  \bibinfo{title}{{M}agnetic behavior of epitaxial {SrRuO$_3$} thin films under
  pressure up to 23 {GP}a}, \bibinfo{journal}{Appl. Phys. Lett.}
  \textbf{\bibinfo{volume}{80}}, \bibinfo{pages}{2338} (\bibinfo{year}{2002}).
\bibAnnoteFile{Lemarrec2002}

\bibitem{Laukhin2012}
\bibinfo{author}{V.~Laukhin}, \bibinfo{author}{O.~Copie},
  \bibinfo{author}{M.~J. Rozenberg}, \bibinfo{author}{R.~Weht},
  \bibinfo{author}{K.~Bouzehouane}, \bibinfo{author}{N.~Reyren},
  \bibinfo{author}{E.~Jacquet}, \bibinfo{author}{M.~Bibes},
  \bibinfo{author}{A.~Barthelemy}, and \bibinfo{author}{G.~Herranz},
  \bibinfo{title}{{E}lectronic subband reconfiguration in a {$d^0$}-perovskite
  induced by strain-driven structural transformations}, \bibinfo{journal}{Phys.
  Rev. Lett.} \textbf{\bibinfo{volume}{109}}, \bibinfo{pages}{226601}
  (\bibinfo{year}{2012}).
\bibAnnoteFile{Laukhin2012}

\bibitem{Hilgenkamp2013}
\bibinfo{author}{H.~Hilgenkamp}, \bibinfo{title}{{N}ovel transport phenomena at
  complex oxide interfaces}, \bibinfo{journal}{MRS Bull.}
  \textbf{\bibinfo{volume}{38}}, \bibinfo{pages}{1026} (\bibinfo{year}{2013}).
\bibAnnoteFile{Hilgenkamp2013}

\bibitem{Thiel2006}
\bibinfo{author}{S.~Thiel}, \bibinfo{author}{G.~Hammerl},
  \bibinfo{author}{A.~Schmehl}, \bibinfo{author}{C.~W. Schneider}, and
  \bibinfo{author}{J.~Mannhart}, \bibinfo{title}{{T}unable
  quasi-two-dimensional electron gases in oxide heterostructures},
  \bibinfo{journal}{Science} \textbf{\bibinfo{volume}{313}},
  \bibinfo{pages}{1942} (\bibinfo{year}{2006}).
\bibAnnoteFile{Thiel2006}

\bibitem{Huijben2009}
\bibinfo{author}{M.~Huijben}, \bibinfo{author}{A.~Brinkman},
  \bibinfo{author}{G.~Koster}, \bibinfo{author}{G.~Rijnders},
  \bibinfo{author}{H.~Hilgenkamp}, and \bibinfo{author}{D.~H.~A. Blank},
  \bibinfo{title}{{S}tructure-{P}roperty {R}elation of {SrTiO$_3$/LaAlO$_3$}
  {I}nterfaces}, \bibinfo{journal}{Adv. Mater.} \textbf{\bibinfo{volume}{21}},
  \bibinfo{pages}{1665} (\bibinfo{year}{2009}).
\bibAnnoteFile{Huijben2009}

\bibitem{Wadati2009}
\bibinfo{author}{H.~Wadati}, \bibinfo{author}{D.~G. Hawthorn},
  \bibinfo{author}{J.~Geck}, \bibinfo{author}{T.~Higuchi},
  \bibinfo{author}{Y.~Hikita}, \bibinfo{author}{H.~Y. Hwang},
  \bibinfo{author}{L.~F. Kourkoutis}, \bibinfo{author}{D.~A. Muller},
  \bibinfo{author}{S.-W. Huang}, \bibinfo{author}{D.~J. Huang},
  \bibinfo{author}{H.-J. Lin}, \bibinfo{author}{C.~Schuessler-Langeheine},
  \bibinfo{author}{H.-H. Wu}, \bibinfo{author}{E.~Schierle},
  \bibinfo{author}{E.~Weschke}, \bibinfo{author}{N.~J.~C. Ingle}, and
  \bibinfo{author}{G.~A. Sawatzky}, \bibinfo{title}{{R}esonant soft x-ray
  scattering studies of interface reconstructions in {SrTiO$_3$/LaAlO$_3$}
  superlattices}, \bibinfo{journal}{J. Appl. Phys.}
  \textbf{\bibinfo{volume}{106}}, \bibinfo{pages}{083705}
  (\bibinfo{year}{2009}).
\bibAnnoteFile{Wadati2009}

\bibitem{Pavlenko2012}
\bibinfo{author}{N.~Pavlenko}, \bibinfo{author}{T.~Kopp},
  \bibinfo{author}{E.~Y. Tsymbal}, \bibinfo{author}{J.~Mannhart}, and
  \bibinfo{author}{G.~A. Sawatzky}, \bibinfo{title}{Oxygen vacancies at
  titanate interfaces: {Two}-dimensional magnetism and orbital reconstruction},
  \bibinfo{journal}{Phys. Rev. B} \textbf{\bibinfo{volume}{86}},
  \bibinfo{pages}{064431} (\bibinfo{year}{2012}).
\bibAnnoteFile{Pavlenko2012}

\bibitem{Popovic2008}
\bibinfo{author}{Z.~S. Popovic}, \bibinfo{author}{S.~Satpathy}, and
  \bibinfo{author}{R.~M. Martin}, \bibinfo{title}{{O}rigin of the
  {T}wo-{D}imensional {E}lectron {G}as {C}arrier {D}ensity at the {LaAlO$_3$}
  on {SrTiO$_3$} {I}nterface}, \bibinfo{journal}{Phys. Rev. Lett.}
  \textbf{\bibinfo{volume}{101}}, \bibinfo{pages}{256801}
  (\bibinfo{year}{2008}).
\bibAnnoteFile{Popovic2008}

\bibitem{Salluzzo2009}
\bibinfo{author}{M.~Salluzzo}, \bibinfo{author}{J.~C. Cezar},
  \bibinfo{author}{N.~B. Brookes}, \bibinfo{author}{V.~Bisogni},
  \bibinfo{author}{G.~M. De~Luca}, \bibinfo{author}{C.~Richter},
  \bibinfo{author}{S.~Thiel}, \bibinfo{author}{J.~Mannhart},
  \bibinfo{author}{M.~Huijben}, \bibinfo{author}{A.~Brinkman},
  \bibinfo{author}{G.~Rijnders}, and \bibinfo{author}{G.~Ghiringhelli},
  \bibinfo{title}{{O}rbital {R}econstruction and the {T}wo-{D}imensional
  {E}lectron {G}as at the {LaAlO$_3$/SrTiO$_3$} {I}nterface},
  \bibinfo{journal}{Phys. Rev. Lett.} \textbf{\bibinfo{volume}{102}},
  \bibinfo{pages}{166804} (\bibinfo{year}{2009}).
\bibAnnoteFile{Salluzzo2009}

\bibitem{Fuchs2015}
\bibinfo{author}{D.~Fuchs}, \bibinfo{author}{A.~Sleem},
  \bibinfo{author}{R.~Sch\"afer}, \bibinfo{author}{A.~G. Zaitsev},
  \bibinfo{author}{M.~Meffert}, \bibinfo{author}{D.~Gerthsen},
  \bibinfo{author}{R.~Schneider}, and \bibinfo{author}{H.~v.~L\"ohneysen},
  \bibinfo{title}{Incipient localization of charge carriers in the
  two-dimensional electron system in LaAlO$_3$/SrTiO$_3$ under hydrostatic
  pressure}, \bibinfo{journal}{Phys. Rev. B} \textbf{\bibinfo{volume}{92}},
  \bibinfo{pages}{155313} (\bibinfo{year}{2015}).
\bibAnnoteFile{Fuchs2015}

\bibitem{Ernst1994}
\bibinfo{author}{S.~Ernst}, \bibinfo{author}{A.~R. Go\~ni},
  \bibinfo{author}{K.~Syassen}, and \bibinfo{author}{K.~Eberl},
  \bibinfo{title}{{C}ollapse of the {H}artree term of the {C}oulomb interaction
  in a very dilute 2{D} electron gas}, \bibinfo{journal}{Phys. Rev. Lett.}
  \textbf{\bibinfo{volume}{72}}, \bibinfo{pages}{4029} (\bibinfo{year}{1994}).
\bibAnnoteFile{Ernst1994}

\bibitem{Schneider2006}
\bibinfo{author}{C.~W. Schneider}, \bibinfo{author}{S.~Thiel},
  \bibinfo{author}{G.~Hammerl}, \bibinfo{author}{C.~Richter}, and
  \bibinfo{author}{J.~Mannhart}, \bibinfo{title}{{M}icrolithography of electron
  gases formed at interfaces in oxide heterostructures},
  \bibinfo{journal}{Appl. Phys. Lett.} \textbf{\bibinfo{volume}{89}}
  (\bibinfo{year}{2006}).
\bibAnnoteFile{Schneider2006}

\bibitem{Altmeyer2015}
\bibinfo{author}{M.~Altmeyer}, \bibinfo{author}{H.~O. Jeschke},
  \bibinfo{author}{O.~Hijano-Cubelos}, \bibinfo{author}{C.~Martins},
  \bibinfo{author}{F.~Lechermann}, \bibinfo{author}{K.~Koepernik},
  \bibinfo{author}{A.~Santander-Syro}, \bibinfo{author}{M.~J. Rozenberg},
  \bibinfo{author}{R.~Valenti}, and \bibinfo{author}{M.~Gabay},
  \bibinfo{title}{Magnetism, spin texture and in-gap states: Atomic
  specialization at the surface of oxygen-deficient SrTiO$_3$},
  \bibinfo{journal}{arXiv preprint arXiv:1511.08614, Phys. Rev. Lett (in
  press)}  (\bibinfo{year}{2015}).
\bibAnnoteFile{Altmeyer2015}

\bibitem{Borisov2015}
\bibinfo{author}{V.~Borisov}, \bibinfo{author}{S.~Ostanin}, and
  \bibinfo{author}{I.~Mertig}, \bibinfo{title}{Two-dimensional electron gas and
  its electric control at the interface between ferroelectric and
  antiferromagnetic insulator studied from first principles},
  \bibinfo{journal}{Phys. Chem. Chem. Phys.} \textbf{\bibinfo{volume}{17}},
  \bibinfo{pages}{12812} (\bibinfo{year}{2015}).
\bibAnnoteFile{Borisov2015}

\bibitem{Bloechl1994a}
\bibinfo{author}{P.~E. Bl{\"o}chl}, \bibinfo{title}{Projector augmented-wave
  method}, \bibinfo{journal}{Phys. Rev. B} \textbf{\bibinfo{volume}{50}},
  \bibinfo{pages}{17953} (\bibinfo{year}{1994}).
\bibAnnoteFile{Bloechl1994a}

\bibitem{Kresse1996}
\bibinfo{author}{G.~Kresse} and \bibinfo{author}{J.~Furthm\"uller},
  \bibinfo{title}{Efficient iterative schemes for \textit{ab initio}
  total-energy calculations using a plane-wave basis set},
  \bibinfo{journal}{Phys. Rev. B} \textbf{\bibinfo{volume}{54}},
  \bibinfo{pages}{11169} (\bibinfo{year}{1996}).
\bibAnnoteFile{Kresse1996}

\bibitem{Hafner2008}
\bibinfo{author}{J.~Hafner}, \bibinfo{title}{{\textit Ab-initio} simulations of
  materials using {VASP}: Density-functional theory and beyond},
  \bibinfo{journal}{J. Comp. Chem.} \textbf{\bibinfo{volume}{29}},
  \bibinfo{pages}{2044} (\bibinfo{year}{2008}).
\bibAnnoteFile{Hafner2008}

\bibitem{Perdew1996}
\bibinfo{author}{J.~P. Perdew}, \bibinfo{author}{K.~Burke}, and
  \bibinfo{author}{M.~Ernzerhof}, \bibinfo{title}{Generalized gradient
  approximation made simple}, \bibinfo{journal}{Phys. Rev. Lett.}
  \textbf{\bibinfo{volume}{77}}, \bibinfo{pages}{3865} (\bibinfo{year}{1996}).
\bibAnnoteFile{Perdew1996}

\bibitem{Dudarev1998}
\bibinfo{author}{S.~L. Dudarev}, \bibinfo{author}{G.~A. Botton},
  \bibinfo{author}{S.~Y. Savrasov}, \bibinfo{author}{C.~J. Humphreys}, and
  \bibinfo{author}{A.~P. Sutton}, \bibinfo{title}{Electron-energy-loss spectra
  and the structural stability of nickel oxide: An {LSDA+$U$} study},
  \bibinfo{journal}{Phys. Rev. B} \textbf{\bibinfo{volume}{57}},
  \bibinfo{pages}{1505} (\bibinfo{year}{1998}).
\bibAnnoteFile{Dudarev1998}

\bibitem{Czyzyk1994}
\bibinfo{author}{M.~T. Czyzyk} and \bibinfo{author}{G.~A. Sawatzky},
  \bibinfo{title}{Local-density functional and on-site correlations: The
  electronic structure of {La$_2$CuO$_4$ and LaCuO$_3$}},
  \bibinfo{journal}{Phys. Rev. B} \textbf{\bibinfo{volume}{49}},
  \bibinfo{pages}{14211} (\bibinfo{year}{1994}).
\bibAnnoteFile{Czyzyk1994}

\bibitem{Bloechl1994b}
\bibinfo{author}{P.~E. Bl{\"o}chl}, \bibinfo{author}{O.~Jepsen}, and
  \bibinfo{author}{O.~K. Andersen}, \bibinfo{title}{Improved tetrahedron method
  for {Brillouin-zone} integrations}, \bibinfo{journal}{Phys. Rev. B}
  \textbf{\bibinfo{volume}{49}}, \bibinfo{pages}{223} (\bibinfo{year}{1994}).
\bibAnnoteFile{Bloechl1994b}

\bibitem{Momma11}
\bibinfo{author}{K.~Momma} and \bibinfo{author}{F.~Izumi},
  \bibinfo{title}{VESTA 3 for three-dimensional visualization of crystal,
  volumetric and morphology data}, \bibinfo{journal}{J. Appl. Cryst.}
  \textbf{\bibinfo{volume}{44}}, \bibinfo{pages}{1272} (\bibinfo{year}{2011}).
\bibAnnoteFile{Momma11}

\bibitem{Hanghui2010}
\bibinfo{author}{H.~Chen}, \bibinfo{author}{A.~M. Kolpak}, and
  \bibinfo{author}{S.~Ismail-Beigi}, \bibinfo{title}{Electronic and Magnetic
  Properties of c Interfaces from First Principles}, \bibinfo{journal}{Adv.
  Mater.} \textbf{\bibinfo{volume}{22}}, \bibinfo{pages}{2881}
  (\bibinfo{year}{2010}).
\bibAnnoteFile{Hanghui2010}

\bibitem{Pentcheva2009}
\bibinfo{author}{R.~Pentcheva} and \bibinfo{author}{W.~E. Pickett},
  \bibinfo{title}{Avoiding the Polarization Catastrophe in LaAlO$_3$ Overlayers
  on SrTiO$_3$ (001) through Polar Distortion}, \bibinfo{journal}{Phys. Rev.
  Lett.} \textbf{\bibinfo{volume}{102}}, \bibinfo{pages}{107602}
  (\bibinfo{year}{2009}).
\bibAnnoteFile{Pentcheva2009}

\bibitem{Pavlenko2011}
\bibinfo{author}{N.~Pavlenko} and \bibinfo{author}{T.~Kopp},
  \bibinfo{title}{Structural relaxation and metal-insulator transition at the
  interface between {SrTiO$_3$ and LaAlO$_3$}}, \bibinfo{journal}{Surf. Sci.}
  \textbf{\bibinfo{volume}{605}} (\bibinfo{year}{2011}).
\bibAnnoteFile{Pavlenko2011}

\bibitem{Piskunov2004}
\bibinfo{author}{S.~Piskunov}, \bibinfo{author}{E.~Heifets},
  \bibinfo{author}{R.~Eglitis}, and \bibinfo{author}{G.~Borstel},
  \bibinfo{title}{Bulk properties and electronic structure of {SrTiO$_3$},
  {BaTiO$_3$}, {PbTiO$_3$} perovskites: an \textit{ab initio} HF/DFT study},
  \bibinfo{journal}{Comput. Mater. Sci.} \textbf{\bibinfo{volume}{29}},
  \bibinfo{pages}{165 } (\bibinfo{year}{2004}).
\bibAnnoteFile{Piskunov2004}

\bibitem{Joshua2012}
\bibinfo{author}{A.~Joshua}, \bibinfo{author}{S.~Pecker},
  \bibinfo{author}{J.~Ruhman}, \bibinfo{author}{E.~Altman}, and
  \bibinfo{author}{S.~Ilani}, \bibinfo{title}{{A} universal critical density
  underlying the physics of electrons at the {LaAlO$_3$/SrTiO$_3$} interface},
  \bibinfo{journal}{Nat. Commun.} \textbf{\bibinfo{volume}{3}}
  (\bibinfo{year}{2012}).
\bibAnnoteFile{Joshua2012}

\bibitem{Xie2013}
\bibinfo{author}{Y.~Xie}, \bibinfo{author}{C.~Bell},
  \bibinfo{author}{Y.~Hikita}, \bibinfo{author}{S.~Harashima}, and
  \bibinfo{author}{H.~Y. Hwang}, \bibinfo{title}{{E}nhancing {E}lectron
  {M}obility at the {LaAlO$_3$/SrTiO$_3$} {I}nterface by {S}urface {C}ontrol},
  \bibinfo{journal}{Adv. Mater.} \textbf{\bibinfo{volume}{25}},
  \bibinfo{pages}{4735} (\bibinfo{year}{2013}).
\bibAnnoteFile{Xie2013}

\bibitem{PyProcar}
\bibinfo{author}{A.~H. Romero} and \bibinfo{author}{F.~Mu\~{n}oz},
  \bibinfo{title}{\textit{PyProcar} code (available at
  https://sourceforge.net/projects/pyprocar/).}  (\bibinfo{year}{2013}).
\bibAnnoteFile{PyProcar}

\bibitem{KingSmith1993}
\bibinfo{author}{R.~D. King-Smith} and \bibinfo{author}{D.~Vanderbilt},
  \bibinfo{title}{Theory of polarization of crystalline solids},
  \bibinfo{journal}{Phys. Rev. B} \textbf{\bibinfo{volume}{47}},
  \bibinfo{pages}{1651} (\bibinfo{year}{1993}).
\bibAnnoteFile{KingSmith1993}

\bibitem{Spaldin2012}
\bibinfo{author}{N.~A. Spaldin}, \bibinfo{title}{A beginner's guide to the
  modern theory of polarization}, \bibinfo{journal}{J. Solid State Chem.}
  \textbf{\bibinfo{volume}{195}}, \bibinfo{pages}{2} (\bibinfo{year}{2012}).
\bibAnnoteFile{Spaldin2012}

\bibitem{Berner2013a}
\bibinfo{author}{G.~Berner}, \bibinfo{author}{A.~M\"uller},
  \bibinfo{author}{F.~Pfaff}, \bibinfo{author}{J.~Walde},
  \bibinfo{author}{C.~Richter}, \bibinfo{author}{J.~Mannhart},
  \bibinfo{author}{S.~Thiess}, \bibinfo{author}{A.~Gloskovskii},
  \bibinfo{author}{W.~Drube}, \bibinfo{author}{M.~Sing}, and
  \bibinfo{author}{R.~Claessen}, \bibinfo{title}{Band alignment in
  {LaAlO$_3$/SrTiO$_3$} oxide heterostructures inferred from hard x-ray
  photoelectron spectroscopy}, \bibinfo{journal}{Phys. Rev. B}
  \textbf{\bibinfo{volume}{88}}, \bibinfo{pages}{115111}
  (\bibinfo{year}{2013}).
\bibAnnoteFile{Berner2013a}

\bibitem{Segal2009}
\bibinfo{author}{Y.~Segal}, \bibinfo{author}{J.~H. Ngai},
  \bibinfo{author}{J.~W. Reiner}, \bibinfo{author}{F.~J. Walker}, and
  \bibinfo{author}{C.~H. Ahn}, \bibinfo{title}{X-ray photoemission studies of
  the metal-insulator transition in {LaAlO$_3$/SrTiO$_3$} structures grown by
  molecular beam epitaxy}, \bibinfo{journal}{Phys. Rev. B}
  \textbf{\bibinfo{volume}{80}}, \bibinfo{pages}{241107}
  (\bibinfo{year}{2009}).
\bibAnnoteFile{Segal2009}

\bibitem{Delugas2005}
\bibinfo{author}{P.~Delugas}, \bibinfo{author}{V.~Fiorentini}, and
  \bibinfo{author}{A.~Filippetti}, \bibinfo{title}{Dielectric properties and
  long-wavelength optical modes of the high-$\ensuremath{\kappa}$ oxide
  LaAlO$_3$}, \bibinfo{journal}{Phys. Rev. B} \textbf{\bibinfo{volume}{71}},
  \bibinfo{pages}{134302} (\bibinfo{year}{2005}).
\bibAnnoteFile{Delugas2005}

\bibitem{Gajdos2006}
\bibinfo{author}{M.~Gajdo\v{s}}, \bibinfo{author}{K.~Hummer},
  \bibinfo{author}{G.~Kresse}, \bibinfo{author}{J.~Furthm\"uller}, and
  \bibinfo{author}{F.~Bechstedt}, \bibinfo{title}{Linear optical properties in
  the projector-augmented wave methodology}, \bibinfo{journal}{Phys. Rev. B}
  \textbf{\bibinfo{volume}{73}}, \bibinfo{pages}{045112}
  (\bibinfo{year}{2006}).
\bibAnnoteFile{Gajdos2006}

\bibitem{Samara1966}
\bibinfo{author}{G.~A. Samara}, \bibinfo{title}{Pressure and Temperature
  Dependences of the Dielectric Properties of the Perovskites {BaTiO$_3$ and
  SrTiO$_3$}}, \bibinfo{journal}{Phys. Rev.} \textbf{\bibinfo{volume}{151}},
  \bibinfo{pages}{378} (\bibinfo{year}{1966}).
\bibAnnoteFile{Samara1966}

\bibitem{Hegenbarth1967}
\bibinfo{author}{E.~Hegenbarth} and \bibinfo{author}{C.~Frenzel},
  \bibinfo{title}{{P}ressure dependence of dielectric constants of {SrTiO$_3$}
  and {Ba$_{0.05}$Sr$_{0.95}$TiO$_3$} At Low Temperatures},
  \bibinfo{journal}{Cryogenics} \textbf{\bibinfo{volume}{7}},
  \bibinfo{pages}{331} (\bibinfo{year}{1967}).
\bibAnnoteFile{Hegenbarth1967}

\bibitem{Martin1975}
\bibinfo{author}{G.~Martin} and \bibinfo{author}{E.~Hegenbarth},
  \bibinfo{title}{{I}nvestigation of pressure effects on solid
  {(Ba$_x$Sr$_{1-x}$TiO$_3$} solutions (x$<$0.1) and {SrTiO$_3$} at low
  temperatures}, \bibinfo{journal}{Journal of Low Temperature Physics}
  \textbf{\bibinfo{volume}{18}}, \bibinfo{pages}{101} (\bibinfo{year}{1975}).
\bibAnnoteFile{Martin1975}

\bibitem{Wang2000a}
\bibinfo{author}{R.~P. Wang}, \bibinfo{author}{N.~Sakamoto}, and
  \bibinfo{author}{M.~Itoh}, \bibinfo{title}{{E}ffects of pressure on the
  dielectric properties of {SrTi$^{18}$O$_3$} and {SrTi$^{16}$O$_3$} single
  crystals}, \bibinfo{journal}{Physical Review B}
  \textbf{\bibinfo{volume}{62}}, \bibinfo{pages}{R3577} (\bibinfo{year}{2000}).
\bibAnnoteFile{Wang2000a}

\bibitem{Yu2014}
\bibinfo{author}{L.~Yu} and \bibinfo{author}{A.~Zunger}, \bibinfo{title}{A
  polarity-induced defect mechanism for conductivity and magnetism at
  polar–nonpolar oxide interfaces}, \bibinfo{journal}{Nat. Commun.}
  \textbf{\bibinfo{volume}{5}}, \bibinfo{pages}{1} (\bibinfo{year}{2014}).
\bibAnnoteFile{Yu2014}

\bibitem{Zhou2015A}
\bibinfo{author}{J.~Zhou}, \bibinfo{author}{T.~C. Asmara},
  \bibinfo{author}{M.~Yang}, \bibinfo{author}{G.~A. Sawatzky},
  \bibinfo{author}{Y.~P. Feng}, and \bibinfo{author}{A.~Rusydi},
  \bibinfo{title}{Interplay of electronic reconstructions, surface oxygen
  vacancies, and lattice distortions in insulator-metal transition of
  {LaAlO$_3$/SrTiO$_3$}}, \bibinfo{journal}{Phys. Rev. B}
  \textbf{\bibinfo{volume}{92}}, \bibinfo{pages}{125423}
  (\bibinfo{year}{2015}).
\bibAnnoteFile{Zhou2015A}

\bibitem{Berner2013b}
\bibinfo{author}{G.~Berner}, \bibinfo{author}{M.~Sing},
  \bibinfo{author}{H.~Fujiwara}, \bibinfo{author}{A.~Yasui},
  \bibinfo{author}{Y.~Saitoh}, \bibinfo{author}{A.~Yamasaki},
  \bibinfo{author}{Y.~Nishitani}, \bibinfo{author}{A.~Sekiyama},
  \bibinfo{author}{N.~Pavlenko}, \bibinfo{author}{T.~Kopp},
  \bibinfo{author}{C.~Richter}, \bibinfo{author}{J.~Mannhart},
  \bibinfo{author}{S.~Suga}, and \bibinfo{author}{R.~Claessen},
  \bibinfo{title}{Direct $k$-Space Mapping of the Electronic Structure in an
  Oxide-Oxide Interface}, \bibinfo{journal}{Phys. Rev. Lett.}
  \textbf{\bibinfo{volume}{110}}, \bibinfo{pages}{247601}
  (\bibinfo{year}{2013}).
\bibAnnoteFile{Berner2013b}

\end{thebibliography}

\end{document}